\documentclass[twocolumn]{aastex63}

\pdfoutput=1

\usepackage{natbib}
\usepackage{color}
\usepackage{bm}
\usepackage{hyperref}
\usepackage{float}
\usepackage{latexsym}
\usepackage{amsmath}
\usepackage{amssymb}
\usepackage{multirow}
\usepackage{rotate}
\usepackage{graphicx}
\usepackage{url}
\usepackage{color}
\usepackage[hyphens]{url}

\submitjournal{ApJ}
\received{August 19th, 2019}
\revised{\today}
\begin{document}

\title{The Canada-France Imaging Survey: Reconstructing the Milky Way Star Formation History from its White Dwarf Population}

\shorttitle{The Galactic Star Formation History}
\shortauthors{Fantin et al.}

\smallskip

\author[0000-0003-3816-3254]{Nicholas J. Fantin}
\correspondingauthor{Nicholas J. Fantin}
\email{nfantin@uvic.ca}
\affil{Department of Physics and Astronomy, University of Victoria, Victoria, BC, V8P 1A1, Canada}
\affil{National Research Council of Canada, Herzberg Astronomy \& Astrophysics Research Centre, 5071 W. Saanich Rd, Victoria, BC, V9E 2E7, Canada}

\author{Patrick C{\^o}t{\'e}}
\affil{National Research Council of Canada, Herzberg Astronomy \& Astrophysics Research Centre, 5071 W. Saanich Rd, Victoria, BC, V9E 2E7, Canada}

\author{Alan W. McConnachie}
\affil{National Research Council of Canada, Herzberg Astronomy \& Astrophysics Research Centre, 5071 W. Saanich Rd, Victoria, BC, V9E 2E7, Canada}

\author{Pierre Bergeron} 
\affil{D{\'e}partement de Physique, Universit{\'e} de Montr{\'e}al, C.P. 6128, Succ. Centre-Ville, Montr{\'e}al, QC H3C 3J7, Canada}

\author{Jean-Charles Cuillandre}
\affil{AIM, CEA, CNRS, Universit{\'e} Paris-Saclay, Universit{\'e} Paris Diderot, Sorbonne Paris Cit{\'e}, Observatoire de Paris, PSL University, F-91191 Gif-sur-Yvette Cedex, France}

\author{Stephen D. J. Gwyn}
\affil{National Research Council of Canada, Herzberg Astronomy \& Astrophysics Research Centre, 5071 W. Saanich Rd, Victoria, BC, V9E 2E7, Canada}

\author{Rodrigo A. Ibata}
\affil{Observatoire Astronomique de Strasbourg, CNRS, UMR 7550, F-67000 Strasbourg, France}

\author{Guillaume F. Thomas}
\affil{National Research Council of Canada, Herzberg Astronomy \& Astrophysics Research Centre, 5071 W. Saanich Rd, Victoria, BC, V9E 2E7, Canada}

\author{Raymond G. Carlberg}
\affil{Department of Astronomy and Astrophysics, University of Toronto, 50 St. George Street, Toronto, ON, M5S 3H4, Canada}

\author{S{\'e}bastien Fabbro}
\affil{National Research Council of Canada, Herzberg Astronomy \& Astrophysics Research Centre, 5071 W. Saanich Rd, Victoria, BC, V9E 2E7, Canada}

\author{Misha Haywood}
\affil{GEPI, Observatoire de Paris, PSL Research University, CNRS,Place Jules Janssen, 92190 Meudon, France}

\author{Ariane Lan\c{c}on}
\affil{Observatoire Astronomique, 11 rue de l'Universit{\'e}, 67000 Strasbourg, France}

\author{Geraint F. Lewis}
\affil{Sydney Institute for Astronomy, School of Physics, A28, The University of Sydney, NSW 2006, Australia}

\author{Khyati Malhan}
\affil{The  Oskar  Klein  Centre  for  Cosmoparticle  Physics,  Department  of Physics,  Stockholm  University,  AlbaNova,  10691  Stockholm,  Sweden}

\author{Nicolas F. Martin}
\affil{Universit{\'e} de Strasbourg, CNRS, Observatoire Astronomique de Strasbourg, UMR 7550, F-67000 Strasbourg, France}
\affil{Max-Planck-Institut f\"{u}r Astronomie, K\"{o}nigstuhl 17, D-69117 Heidelberg, Germany}

\author{Julio F. Navarro}
\affil{Department of Physics and Astronomy, University of Victoria, Victoria, BC, V8P 1A1, Canada}

\author{Douglas Scott}
\affil{Department of Physics and Astronomy, University of British Columbia, Vancouver, BC V6T 1Z1, Canada}

\author{Else Starkenburg}
\affil{Leibniz Institute for Astrophysics Potsdam (AIP), An der Sternwarte 16, D-14482 Potsdam, Germany}

\begin{abstract}

As the remnants of stars with initial masses $\lesssim$ 8\,M$_{\odot}$, white dwarfs contain valuable information on the formation histories of stellar populations. In this paper, we use deep, high-quality, \textit{u}-band photometry from the Canada France Imaging Survey (CFIS), \textit{griz} photometry from Pan-STARRS 1 (PS1), as well as proper motions from \textit{Gaia}  DR2, to select 25,156 white dwarf candidates over $\sim$4500\,deg$^2$ using a reduced proper motion diagram. We develop a new white dwarf population synthesis code that returns mock observations of the Galactic field white dwarf population for a given star formation history, while simultaneously taking into account the geometry of the Milky Way, survey parameters, and selection effects. We use this model to derive the star formation histories of the thin disk, thick disk, and stellar halo. Our results show that the Milky Way disk began forming stars (11.3 $\pm$ 0.5)\,Gyr ago, with a peak rate of (8.8 $\pm$ 1.4)\,M$_{\odot}~$yr$^{-1}$ at (9.8 $\pm$ 0.4)\,Gyr, before a slow decline to a constant rate until the present day --- consistent with recent results suggesting a merging event with a satellite galaxy. Studying the residuals between the data and best-fit model shows evidence for a slight increase in star formation over the past 3\,Gyr. We fit the local fraction of helium-atmosphere white dwarfs to be (21 $\pm$ 3)\,\%. Incorporating this methodology with data from future wide-field surveys such as LSST, \textit{Euclid}, \textit{CASTOR}, and \textit{WFIRST} should provide an unprecedented view into the formation of the Milky Way at its earliest epoch through its white dwarfs.

\bigskip

\end{abstract}


\section{Introduction}
\label{sec:intro}

The formation and evolution of disk galaxies, and the Milky Way (MW) in particular, has long been an important topic in astronomy since these galaxies dominate star formation activity in the low-redshift Universe. Previous investigations into the formation and evolution of MW-like galaxies have focused on the assembled mass as a function of redshift \citep[see, e.g,][]{vDokkum2013}, or compared resolved stellar populations to theoretical isochrones in local galaxies such as Andromeda \citep{Ferguson2005,Brown2006}. Within the MW itself, much of the attention has been on the stellar metallicity distribution since this, once combined with a model for the gas infall rate, contains information on the star formation rate (SFR) as a function of time \citep[see, e.g,][]{Snaith2015,Haywood2018,Toyouchi2018}. Results from such Galactic chemical evolution (GCE) models typically show a strong period of initial star formation followed by a slowly decaying trend towards the present day, features usually associated with a thick and thin disk, respectively.

\begin{figure*}[!t]
	
	\includegraphics[angle=0,width=.5\textwidth]{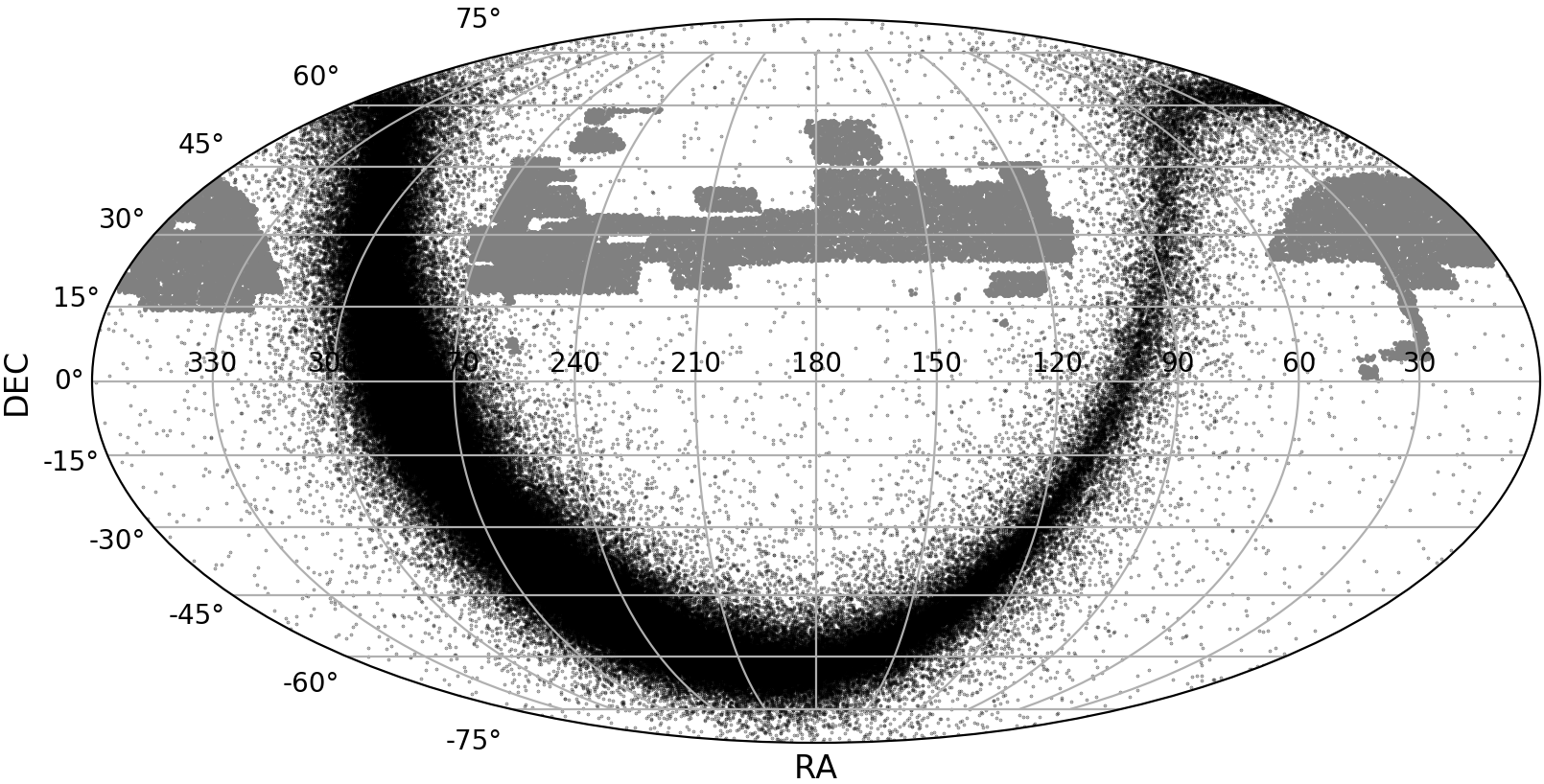}
	\includegraphics[angle=0,width=.5\textwidth]{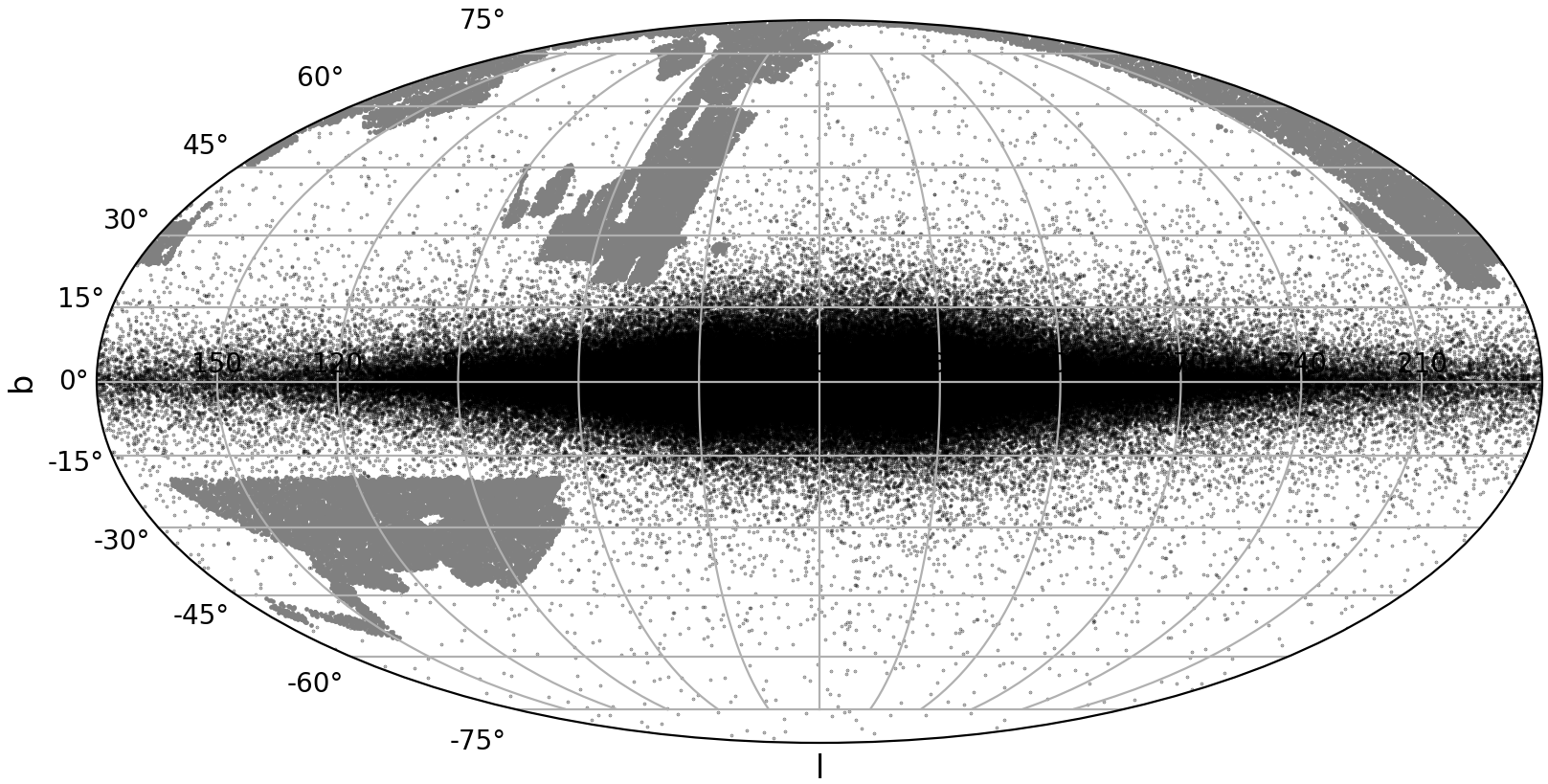}
	\caption{ \textit{Left:} Equatorial positions of thin disk stars within our model. \textit{Right:} The model in Galactic coordinates. In both panels, the shaded region represents the CFIS-\textit{u} footprint as of the end of the 2018A semester. \bigskip }
	\label{fig:Density}
\end{figure*}

The discovery of the dichotomy of the disk began with \cite{Gilmore1983} who used star counts towards the South Galactic Pole and found that the vertical distribution was well described by two separate exponential profiles with different scale heights. Follow-up studies have shown that the thick disk stars are more alpha-rich, iron-poor, and have a larger velocity dispersion --- evidence that they formed much earlier than their thin disk counterparts  \citep{Norris1985,Eggen1998}. Within the past decade, a renewed emphasis has been placed on the early formation of the MW's disk. The disk dichotomy has been questioned by \cite{Bovy2012} who used mono-abundance populations to show that each population could be described by a single exponential scale height. \cite{Haywood2016}, however, argued that the dip in the [$\alpha$/Fe] distribution seen in APOGEE can only be produced by two separate epochs of star formation, and associates each epoch as belonging to the thin and thick disk respectively \citep{Snaith2015, Haywood2018}. \cite{Haywood2016} further argues that these two results are consistent if one assumes that the scale height of the thick disk decreases as a function of time so that the differences in structural parameters between the young thick disk and old thin disk are minor. Such a decrease in the scale height, from 800\,pc to 340\,pc, was found to match star counts in SDSS and 2MASS by \cite{Robin2014}, with a mean scale height of 535 $\pm$ 50\,pc. Further emphasis has been placed on this transitional epoch with recent results by \cite{Helmi2018} and \cite{Belokurov2018}, who show that this epoch of the MW may have played host to a merger with at least one satellite galaxy, a scenario that would likely play an important role in the formation of the thick disk.

With a complete star formation history, the stellar mass of each component can be estimated. Traditionally, star count models calibrated to the solar neighborhood have found that the thick disk contributes about 12\% of the local mass \citep{Juric2008}. However, a recent study by \cite{Snaith2015} using the chemical abundances of local F, G, and K-type stars concluded that the thin and thick disk masses may be comparable. This discrepancy likely arises due to the regions of the MW which are probed by each survey. Many models use data confined to the solar neighborhood, a location at which the thin disk dominates, and thus extrapolating local results to the entire disk becomes problematic.

While the star formation history is imprinted in the chemical enrichment of the MW, stars also leave behind another important tool for astronomers, white dwarfs. White dwarfs form at the end of the asymptotic giant phase as the outer layers are ejected to expose the degenerate carbon-oxygen core. Since these stars are remnants of stars with initial masses less than 8 M$_{\odot}$, they contain valuable information regarding the formation history of stellar populations as a whole.

Since their evolution is well understood --- the age is primarily a function of temperature, luminosity, and atmospheric composition --- white dwarfs have been used in the past to estimate ages. These techniques have been applied to various stellar populations, from globular clusters \citep[see, e.g,][]{Richer2006,Hansen2007,Bedin2009} to MW components \citep{Kilic2017}. The first such study of the MW's disk was carried out by \cite{Winget1987} who used the Luyten Half-Second proper motion catalog to estimate an age of (9.3 $\pm$ 2.0)\,Gyr for the Galactic disk, and (10.3 $\pm$ 2.2)\,Gyr for the age of the Universe, where the age refers to the onset of star formation. These early studies, however, suffered from strong incompleteness and small number statistics due to shallow photometric data. A more recent analysis by \cite{Kilic2017} used the white dwarf luminosity function from \cite{Munn2017} to simultaneously derive the ages of the MW components, obtaining age measurements of (7.8 $\pm$ 0.4)\,Gyr for the thin disk and (9.7 $\pm$ 0.2)\,Gyr for the thick disk. Due to the small number of halo white dwarfs in their sample, the age of the halo was essentially unconstrained. 


With a large enough sample, the complete star formation history of a stellar population can be determined from white dwarfs in two ways. First, with accurate distances one can calculate the luminosity, radius, and effective temperature and convert this into a cooling age \citep{Giammichele2012}. The total age of a white dwarf can be found by adding the cooling age to the progenitor age, which is found by converting the white dwarf mass to a progenitor mass using the initial-to-final mass relation (IFMR). \cite{Tremblay2014} applied this method to the local (20\,pc) sample and measured the star formation history in the solar neighborhood. Their results suggest a roughly constant SFR over the past 10\,Gyr, with a slight drop 5\,Gyr ago (see their Figure 9). This method, however, is typically constrained to small volumes given the distance uncertainties.

The second method is to invert the white dwarf luminosity function (WDLF). \cite{Noh1990} showed that varying the functional form for the star formation rate would leave imprints on the resulting WDLF, such as the slope and turn-off. This method was used by \cite{Rowell2013} to invert the SDSS WDLF assembled by \cite{Harris2006}, and showed a resulting star formation history composed of two broad peaks (at 2 and 9\,Gyr) separated by a significant dip. The advantage of using this method is that the volume that can be probed is considerably larger; the downside, however, is that volume and completeness corrections must be applied. The recent advances made in wide-field astronomy present an opportunity to increase the sampled volume, and the combination of deep, multi-band photometry and accurate proper motions are ideal for selecting white dwarfs. 

In this paper, we introduce a third technique: forward modeling. This involves simulating a sample of white dwarfs given a star formation prescription. The advantage of forward modeling is that complex systematics, like completeness, observational uncertainties, selection effects, and the structure of the MW can all be accounted for. We use data from the Canada-France Imaging Survey, Pan-STARRS 1, and \textit{Gaia} DR2 to derive the SFH of the MW components using a new population synthesis code. Section \ref{sec:data} describes the dataset, while Section \ref{sec:model} describes our white dwarf population synthesis code. We present the results  in Section \ref{sec:Results}, discuss their implications in Section \ref{sec:Discussion}, and summarize in Section \ref{sec:Conclusion}.

\bigskip
\section{Data}
\label{sec:data}

\begin{figure*}[!t]
	
	\includegraphics[angle=0,width=.5\textwidth]{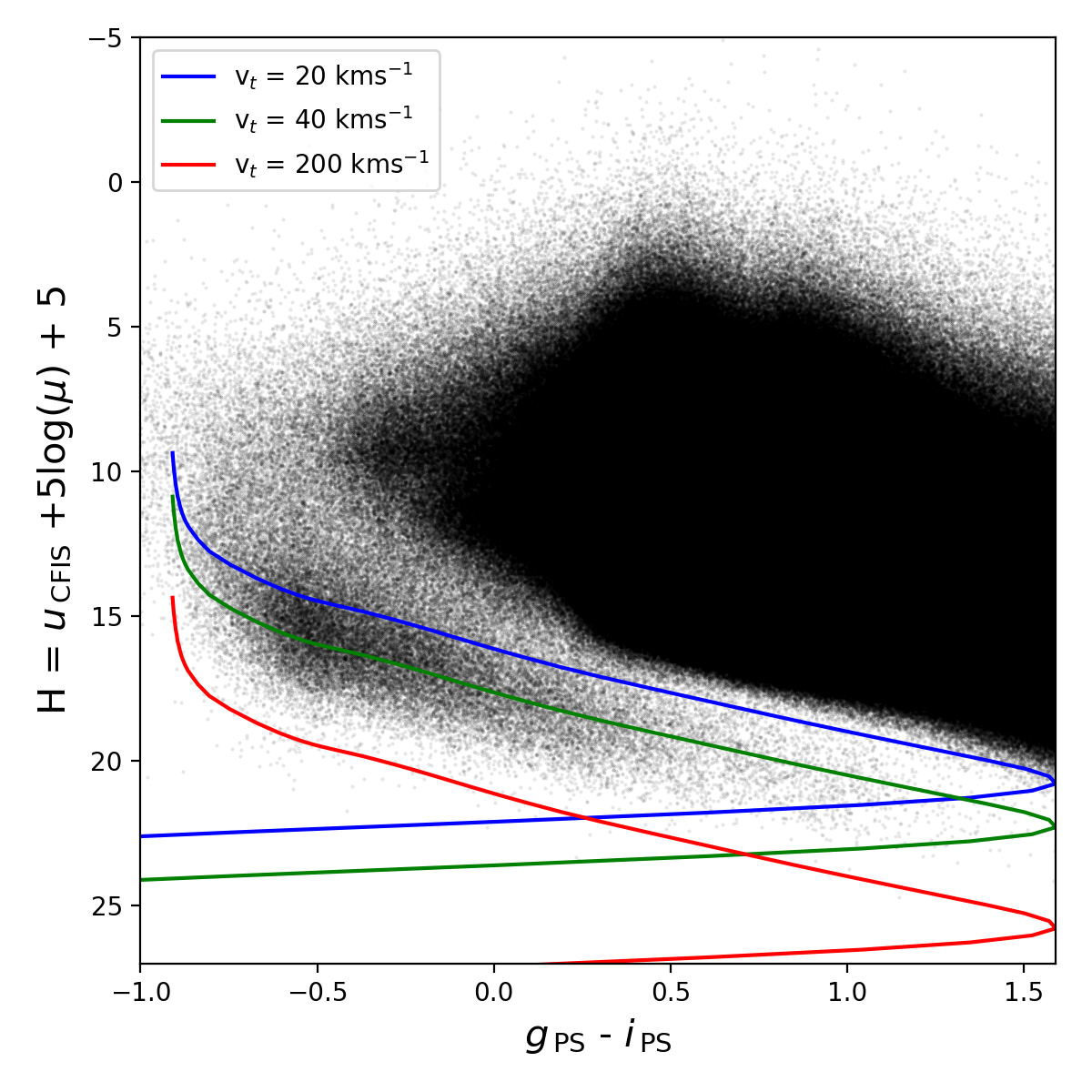}
	\includegraphics[angle=0,width=.5\textwidth]{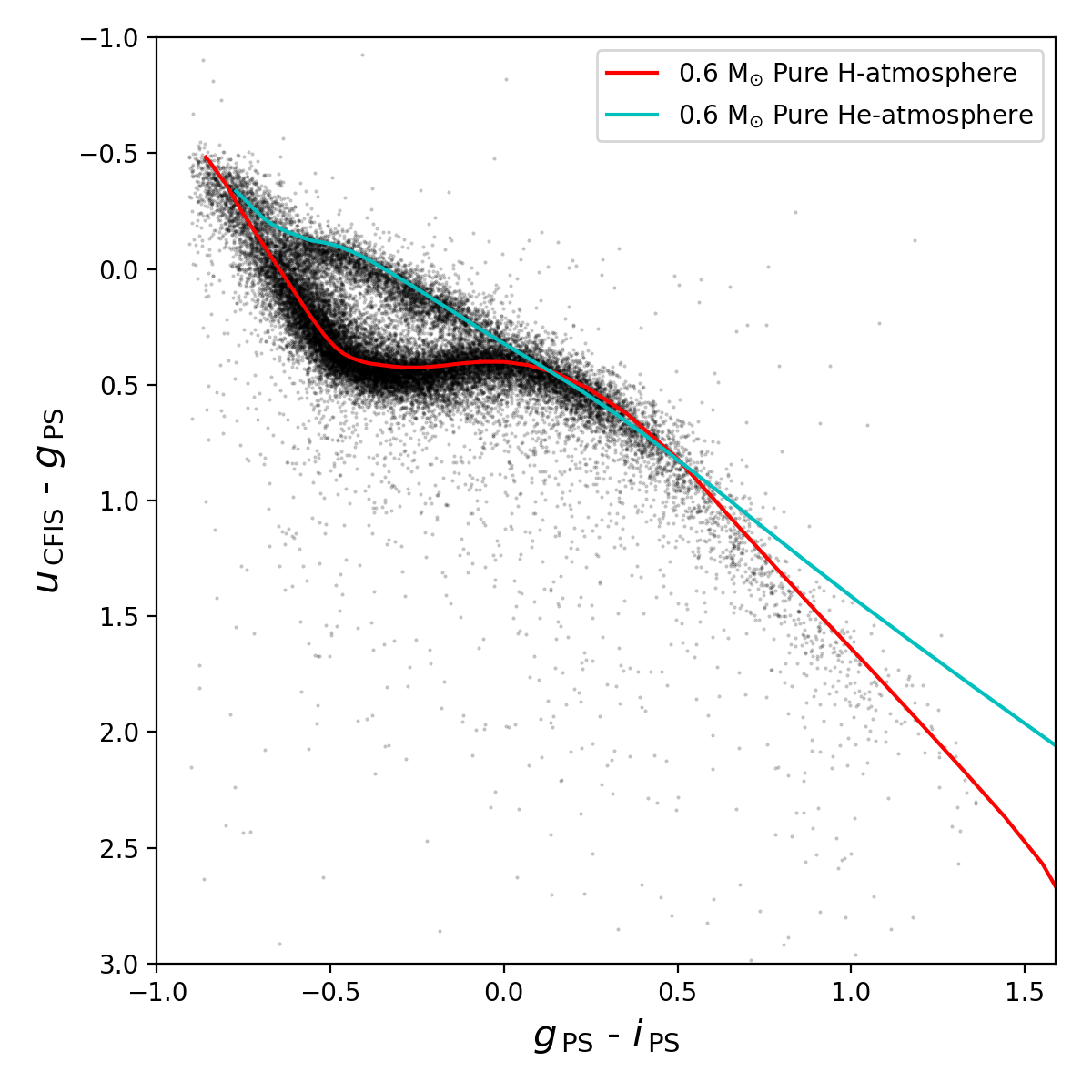}
	\caption{\textit{Left:} CFIS-PS1-\textit{Gaia}  reduced proper motion diagram (RPMD) showing all 24.5 million sources. White dwarfs for this study were selected if they had inferred tangential velocities greater than 20\,kms$^{-1}$ (the region bound by the blue line) \textit{Right:} Color-color diagram showing the resulting white dwarf candidates selected from the RPMD. For reference, 0.6\,M$_{\odot}$ H- and He-atmosphere model tracks are plotted in red and cyan, respectively.
		\bigskip }
	\label{fig:RPMD}
\end{figure*}

\begin{figure*}[!t]
	\centering
	\includegraphics[angle=0,width=0.8\textwidth]{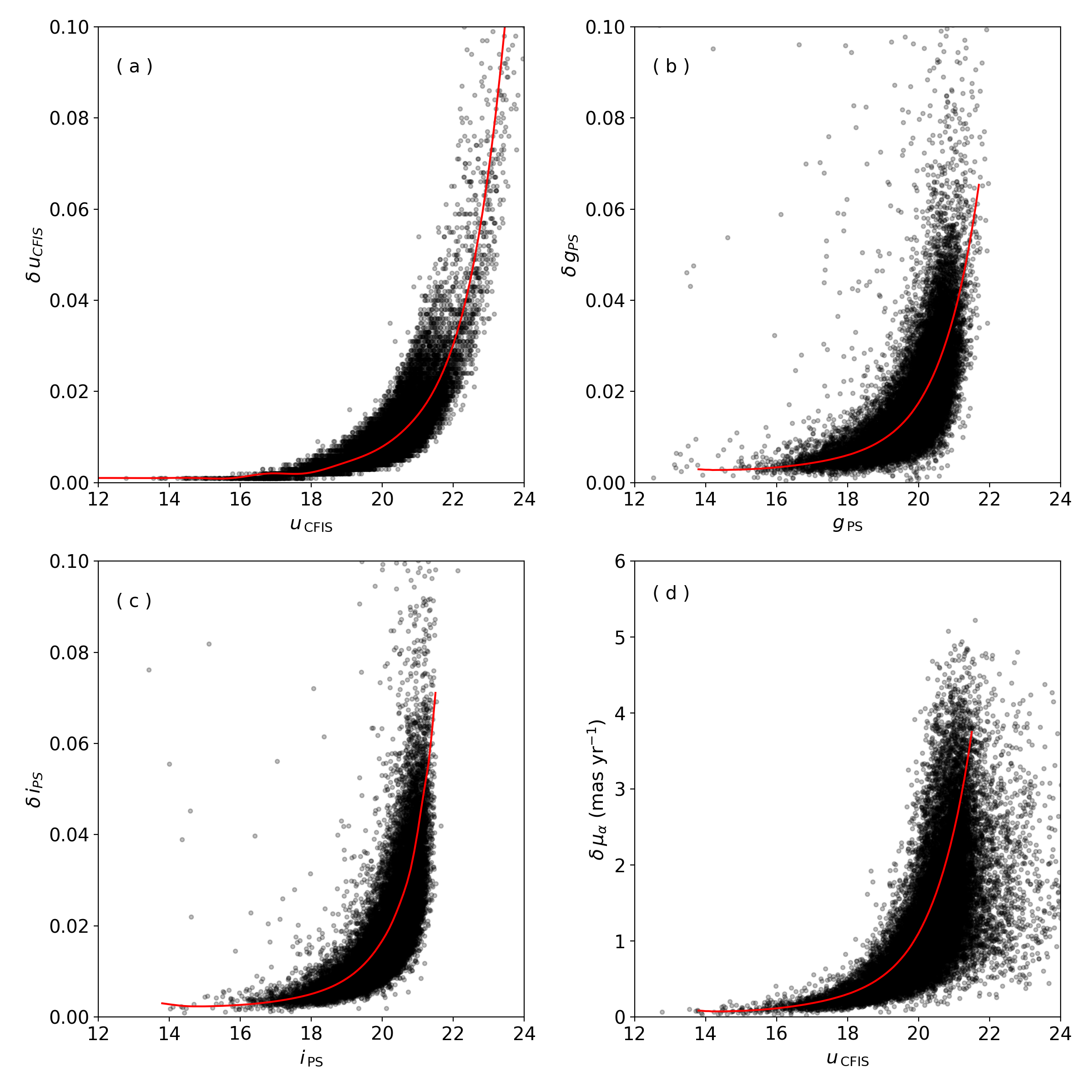}
	\caption{  Observational uncertainties in (a) CFIS \textit{u}, (b) PS1 \textit{g}, (c) PS1 \textit{i}, and (d) proper motions as a function of magnitude in our white dwarf sample. The model, as described in Section \ref{sec:model}, samples a Gaussian at each point with a mean (given by the red line showing a polynomial fit) and standard deviation in order to obtain an uncertainty value for each mock white dwarf. \bigskip }
	\label{fig:Errors}
\end{figure*}

The photometric data used in this paper were acquired as part of the Canada France Imaging Survey \citep[CFIS;][]{CFIS1} and the \textit{grizy} Pan-STARRS 1 3$\pi$ survey \citep[PS1;][]{PS1}. CFIS is an ongoing large-program at the Canada France Hawaii Telescope (CFHT) that aims to obtain 10,000\,deg$^2$ of \textit{u}- and 5,000\,deg$^2$ of \textit{r}-band photometry to 5$\sigma$ point source depths of 24.2 and 24.85 AB mag, respectively. The survey is performed under excellent conditions, with a median \textit{u}-band image quality of 0.$^{\prime\prime}$78. This paper makes use of the CFIS-\textit{u} catalog, which covers $\sim$4,500\,deg$^2$ as of the end of the 2018A semester. The footprint is highlighted in gray, in both equatorial and Galactic coordinates, as part of Figure \ref{fig:Density}.

The CFIS-\textit{u} catalog was merged with PS1 \textit{grizy} forced photometry as described in \cite{Thomas2018}. Given that the PS1 and \textit{Gaia}  catalogs cover the whole sky as seen from CFHT, our area is set by the CFIS-\textit{u} footprint. This catalog retains more than 98\% of PS1 detections within the magnitude range of interest. Stars are selected using the recommended star-galaxy classification from \cite{Farrow2014}: \textit{i}$_{\textrm{PSF}}$ - \textit{i}$_{\textrm{Kron}}$ $<$ 0.05~mag.

We also merge our catalog with \textit{Gaia}  DR2 \citep{Gaia} in order to obtain proper motions. We apply the same astrometric quality cuts on the data as presented in \cite{Gentile2019},  

\begin{align}
\label{equation:Gaia}
\textsc{astrometric\_sigma5d\_max}&<1.5 ~ \textsc{or}\\
(\textsc{astrometric\_excess\_noise}&<1 \nonumber\\
\textsc{and}~ \textsc{parallax\_over\_error}&>4 \nonumber\\
\textsc{and}~\textsc{sqrt}(\textsc{pmra}^{2}+\textsc{pmdec}^{2})&>10\,\textrm{mas})\nonumber,
\end{align}

which removes objects with poor astrometric measurements. Specifically, \textsc{astrometric\_sigma5d\_max} represents the longest axis of the 5-d error ellipsoid, in mas, and large values indicate that one of the astrometric parameters is poorly constrained. \textsc{astrometric\_excess\_noise} is a measure of the difference between the best-fit astrometric solution and the data, with large values suggesting a poor astrometric solution. Finally, \textsc{parallax\_over\_error} is the relation between the parallax measurement and its associated error. The limiting magnitude of \textit{Gaia}  (\textit{G} = 20.7) represents the primary limitation of our catalog, and the implications of this are discussed in Section \ref{sec:Discussion}. Our CFIS-PS1-\textit{Gaia}  catalog contains more than 21.5 million unique point-sources down to the \textit{Gaia} limiting magnitude.

\subsection{White Dwarf Selection}

Using our combined catalog, we have constructed a reduced proper motion diagram (RPMD) based on photometry from CFIS and PS1, and proper motions from \textit{Gaia}  DR2, to separate white dwarfs from other point sources. The reduced proper motion, \textit{H} \citep{Luyten1922}, combines the apparent magnitude, \textit{m}, and the proper motion, \textit{$\mu$}, and can be used as a proxy for absolute magnitude, \textit{M}, and tangential velocity, \textit{v$_{\mathrm{t}}$}:

\begin{equation}
\label{equation:RPM}
\begin{array}{lcl}
H & = & m + 5\log\mu + 5 \\
& = & M + 5\log v_{\mathrm{t}} - 3.379. \\
\end{array}
\end{equation}

Owing to their intrinsic faintness, for a given temperature white dwarfs are well separated in the RPMD from main-sequence stars with similar colors \citep[see, e.g,][]{Harris2006,RowellHambly2011,Fantin2017,Munn2017}. Using model cooling tracks we select white dwarfs with tangential velocities greater than 20\,kms$^{-1}$ as shown in the left-hand panel of Figure \ref{fig:RPMD}. These synthetic magnitudes have been calculated in the CFIS-\textit{u} and PS1 \textit{grizy} bands for H- and He-atmosphere
models using the procedure described in \citet{HolbergBergeron2006}. The white dwarf cooling sequences are similar to those
described in \citet{Fontaine2001} with (50/50) C/O-core compositions, $M_{\rm He}/M_{\star}=10^{-2}$, and $M_{\rm H}/M_{\star}=10^{-4}$ or $10^{-10}$ for H- and He-atmosphere white dwarfs, respectively \footnote{See \url{http://www.astro.umontreal.ca/~bergeron/CoolingModels}}. Our selection procedure identifies 25,156 white dwarf candidates within the CFIS-\textit{u} footprint. The resulting sample, along with model H- and He-atmosphere tracks, can be seen in color-color space in the right-hand panel of Figure \ref{fig:RPMD}.

Previous selections using the RPMD have obtained contamination fractions from non-white dwarfs (typically hot sdO, sdB, or cooler main-sequence stars) of a few percent \citep[e.g,][]{Harris2006} at a velocity cut of $v_{\textrm{t}} > 30\,$kms$^{-1}$, with the rate increasing for lower velocity cuts \citep{Kilic2010}. Given that \textit{Gaia} DR2 has much better astrometry than previous studies, we have chosen to lower our velocity cut to $v_{\textrm{t}} > 20\,$kms$^{-1}$. Matching our objects to SIMBAD, we find 3688 spectroscopic matches, of which 3589 are white dwarfs. The remaining objects are a mix of hot subdwarfs (69 objects), QSOs with spurious proper motion measurements (20 objects), and 10 miscellaneous objects including main-sequence stars. The resulting contamination rate is thus at least $\sim$3\%, a value which has been incorporated into the results presented in Section \ref{sec:Results}.

The color-color diagram highlights the precise photometry of CFIS and PS1 and the ability to separate DA (H-lines present) and DB (He-lines present) white dwarfs between $-$0.6 $<$ \textit{g$_{\textrm{PS}}$} $-$ \textit{i$_{\textrm{PS}}$} $<$ 0.0 mag. The RPMD also highlights the precise astrometry of \textit{Gaia} DR2, which results in a clean sequence of white dwarfs in color-color space. In Figure \ref{fig:Errors} we present the observational uncertainties in both the photometry and astrometry as a function of magnitude for our white dwarf sample, highlighting the depth of the CFIS-\textit{u} band and the precision of the \textit{Gaia} DR2 proper motions.

\subsection{Completeness}

\begin{figure}[!t]
	
	\includegraphics[angle=0,width=.5\textwidth]{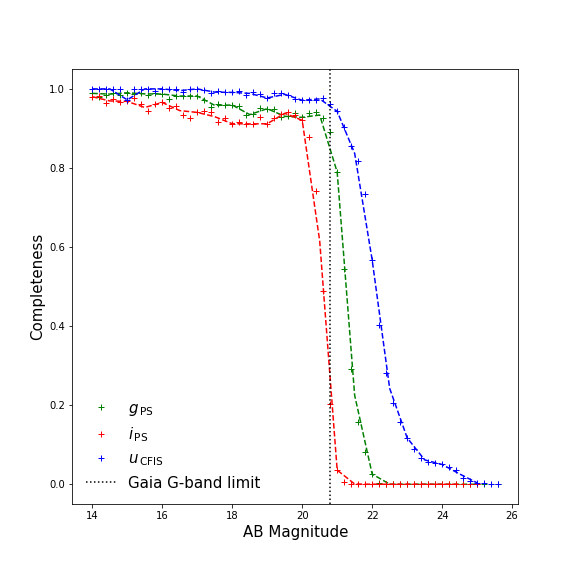}
	\caption{The resulting completeness as a function of magnitude for our CFIS-PS1-\textit{Gaia} sample in the CFIS \textit{u}- (blue), PS1 \textit{g}- (green), and PS1 \textit{i}-band (red) respectively. These completeness functions were calculated assuming that the CFIS-PS1 catalog is complete over the magnitude range of \textit{Gaia} DR2, which was shown to be true in \cite{Thomas2018}.}
	\label{fig:completeness}
\end{figure}

We calculate the completeness within our dataset by comparing our CFIS-PS1-\textit{Gaia} catalog to the CFIS-PS1 catalog. \cite{Thomas2018} showed the PS1 g-band is complete to 22.0 mag, which is deeper than our sample (see Figure \ref{fig:Errors}) and thus we assume that the CFIS-PS1 dataset is complete over the magnitude range covered by the \textit{Gaia} survey. Our completeness curves for the PS1 \textit{g}- and \textit{i}-bands, as well as the CFIS-\textit{u} band, can be seen in Figure \ref{fig:completeness}.

\bigskip
\section{Modeling the Milky Way's White Dwarf Population}
\label{sec:model}

While the MW's white dwarf population contains important information regarding its evolutionary history, it is generally not the focus of stellar population synthesis codes. Prominent population synthesis codes, like the Besan\c{c}on model \citep{Robin2003, Robin2014}, add white dwarfs given an observed local density, while Trilegal \citep{Girardi2005} has issues reproducing observations of white dwarfs \citep{Fantin2017}. Thus, a model that allows white dwarfs to form via the evolution of progenitor stars, as opposed to being added manually, allows one to study various input parameters, such as the star formation history, which can affect the resulting white dwarf population. In this section, we describe our white dwarf population synthesis code and detail our various input assumptions about the MW.

\begin{table*}[!t]
\centering
\caption{Assumed Model Distributions for Each Component}
\label{table:distributions}
\begin{tabular}{ccccccccc}

	\hline \hline
	Component &  $\rho(R,z)$ & $\xi$(M) & SFH & [Fe/H] &$\langle$ V$_{\phi}$ $\rangle$  & $\sigma_{U}$ & $\sigma_{V}$  & $\sigma_{W}$ \\ 
	&&&&&(kms$^{-1}$)&(kms$^{-1}$)&(kms$^{-1}$)&(kms$^{-1}$)\bigskip \\
	\hline
	Thin disk  &  $e^{- R/h_R}e^{- z/h_z}$      & Kroupa &  Skewed Gaussian & 0.0 &  $-$12     & 33 & 15  & 15  \\
	
	&  h$_R$ = 2300\,pc, h$_z$ = 300\,pc      &  && &&& \bigskip\\

	Thick disk   &    $e^{- R/h_R}e^{- z/h_z}$        & Kroupa &  Skewed Gaussian & $-$0.7 &  $-$85      & 40 & 32  & 28  \\
	
	&  h$_R$ = 2300\,pc, h$_z$ = 550\,pc      &  & &&&& \bigskip\\
	
	Stellar halo   &    $r^{-2.44}$      &  Kroupa &   Skewed Gaussian & $-$1.5 &$-$226    & 131  & 106   & 85 \\
	
	\hline
\end{tabular}
\tablecomments{The Solar position is assumed to be ($R_{\odot}$, $\theta_{\odot}$, $z_{\odot}$) = (8340\,pc, 0, 17\,pc) \bigskip}
	
\end{table*}

\subsection{Model Functions}
\bigskip
We construct a synthetic model of the MW's white dwarf population in order to compare to our dataset. We start by assuming a three-component Galaxy consisting of a thin disk, a thick disk, and a stellar halo, with density distributions given in Table \ref{table:distributions}. We assume a double exponential profile in both \textit{R} and \textit{z} for the thin and thick disk. We assume a scale length of 2.3\,kpc for both the thin and thick disk, and scale heights of 300 and 550\,pc respectively, similar to the mean values presented by \cite{Robin2014}. For each component, we generate stars at positions and masses given by the assumed density distributions and \cite{Kroupa2001} initial-mass function (IMF).

Stars are spawned at Galactocentric radii (\textit{R}) and heights above the plane (\textit{z}) appropriate for the distributions shown in Table \ref{table:distributions}.  The stars are evenly distributed in the angular coordinate, with 0$^{\circ}$ representing the line connecting the solar position in the plane to the Galactic center. The Galactocentric coordinates are then converted to RA and Dec, and a distance from the Sun is calculated assuming the Sun is at a position of (\textit{R}, $\theta$, \textit{z}) = (8340\,pc, 0\,pc, 17\,pc) \citep{Reid2014, Karim2017}. Figure \ref{fig:Density} shows the resulting equatorial (left) and Galactic (right) positions for the thin disk with the CFIS-\textit{u} footprint highlighted in gray.

The amount of time it will take each star to become a white dwarf, the progenitor lifetime, is calculated using the analytic stellar lifetimes from \cite{Hurley2000}. This functional form takes the initial mass and the metallicity of each star and returns the lifetime of the star. For our model, we assume solar metallicity for the thin disk, [Fe/H] = $-$0.7 for the thick disk, and [Fe/H] = $-$1.5 for the halo \citep{Peng2013}. Each star is also assigned a birth date (its formation age) that is randomly generated given the functional form for the SFR. We assume a skewed Gaussian star formation history for each component,

\begin{equation}
\label{equation:skew}
\textrm{SFR}(t) = \rho_{0}\frac{2}{\sigma_t}\phi\left ( \frac{t - \xi }{\sigma_t} \right )\Phi\left ( \alpha\left ( \frac{t - \xi }{\sigma_t} \right ) \right )
\end{equation}

\noindent where $\phi(t)$ is the standard normal probability distribution that is symmetric about the mean, $\xi$, and $\Phi(t)$ is its cumulative distribution function. The skewness parameter is $\alpha$, the standard deviation is given by $\sigma_t$, and $\rho_{0}$ is the space density.

\begin{figure*}[!t]
	
ˇ	\includegraphics[angle=0,width=.99\textwidth]{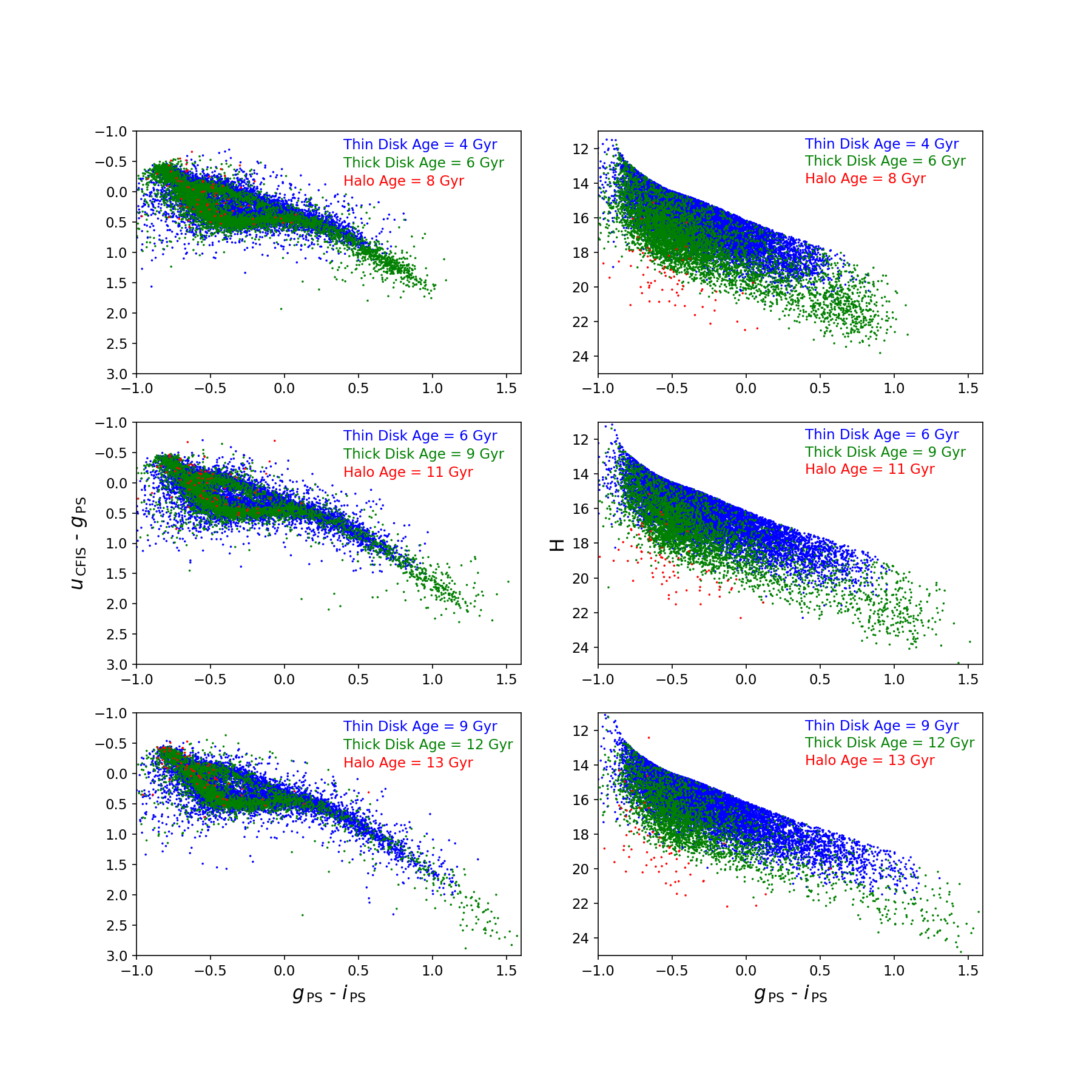}
	\caption{ Comparing three different realizations of the model with young ages (top row), intermediate ages (middle row), and old ages (bottom row) using color-color diagrams (left column) and reduced proper motion diagrams (right column). As the ages of each component increase, a larger number of white dwarfs at redder colors are visible due to the increased cooling ages, resulting in lower temperatures. 
		\bigskip }
	\label{fig:Model}
\end{figure*}

A star will have formed a white dwarf if the lifetime up to the end of the AGB phase is less than its formation age, with this difference being the white dwarf cooling time. The white dwarfs are then given space velocities according to the distributions presented in Table \ref{table:distributions}. These solar position values were adopted from \cite{Robin2017}, who derived them by comparing the Besan\c{c}on model to kinematic data from RAVE and \textit{Gaia} DR1. These values are similar to those derived by \cite{RowellHambly2011} using white dwarfs in the SuperCOSMOS survey. We do not allow the velocities to vary in \textit{R} or \textit{z}, as shown in \cite{Robin2017}, since our sample is local (see Section \ref{sec:Results}) and therefore this gradient would have a minimal impact on our results. We also do not allow our thin disk velocity dispersion to vary with age, but instead take a mean value since the small change would have little effect on a star's position in the RPMD. These velocities are converted to a proper motion using the equations from \cite{JohnsonSoderblom1987}. 

The white dwarf mass, in solar masses, is calculated from the initial progenitor mass using the IFMR from \cite{Kalirai2008}.

\begin{equation}
M_{\textrm{WD}} = (0.109 \pm 0.007)M_{i} + (0.394 \pm 0.025) \textrm{M}_{\odot}.
\end{equation}

\bigskip

The final parameter needed to calculate the photometry of a white dwarf is its spectral type. We include pure H- and He-atmosphere white dwarfs in our model, and each white dwarf is designated as one or the other given an input fraction, $f_{\textrm{He}}$. With the cooling age, white dwarf mass, and type in hand, we determine the absolute magnitudes in the required bands using the white dwarf cooling models as described in Section \ref{sec:data} and seen in Figure \ref{fig:RPMD}. We then determine the reddening in each band at the white dwarfs position using the extinction coefficients and E(B$-$V) values from the Bayestar 3D dust maps of \citep{DustMaps, DustMaps2}, before converting the model absolute magnitudes to apparent magnitudes. We then add experimental uncertainties based on the observed relation between magnitude and error in the given photometric bands and proper motion (see Figure \ref{fig:Errors}). Finally, completeness (see Figure \ref{fig:completeness}) and selection effects (see Figure \ref{fig:RPMD}) are applied based on the observations.

The model returns a catalog of simulated white dwarfs within a given footprint. Catalog information includes positions, distances, masses, cooling ages, spectral types, proper motions, and apparent magnitudes in the given photometric bands. Typical results can be seen in Figure \ref{fig:Model}, which shows color-color diagrams (left column) and RPMDs (right column) for young (top row), intermediate-age (middle row) and old populations (bottom row).

While the above description has been tailored for this particular study, we note that the model can readily be adapted for any future studies by modifying the band-passes and survey parameters. 

\bigskip
\subsection{Fitting Method: Approximate Bayesian Computation MCMC}
\label{sec:calibration}

We begin by parameterizing the SFH for each Galactic component following equation \ref{equation:skew}. This results in four parameters for each component: the mean, $\xi$, which we will call the functional age, the scale, $\sigma_t$, the skewness $\alpha$, and the star formation rate. This results in a total of 12 parameters. We also fit for the ratio between the hydrogen and helium atmosphere white dwarfs, which we call the He fraction, $f_{\textrm{He}}$. This fraction is assumed to be equal for all three populations. This brings the total number of parameters to 13. 

We integrate a given SFH to determine the total stellar mass and select the corresponding number of stars contained in each MW component. For each star, we then generate a formation age given the SFH. We then remove every star that does not fall within the CFIS footprint and/or will not have formed a white dwarf at the present day. The cooling age for those stars that remain is then calculated as being the difference between the formation age and the progenitor lifetime.

With the cooling ages, white dwarf masses, and spectral type, we then calculate apparent magnitudes in the CFIS and PS1 bands using the model cooling curves described in Section \ref{sec:data}. Observational uncertainties are added given the relations shown in Figure \ref{fig:Errors}. A photometric completeness correction is then made given the curves presented in Figure \ref{fig:completeness}.

Finally, we perform the same selection methods on the model as we do on the data. We begin by applying identical distributions in the \textit{Gaia} parameters included in equation \ref{equation:Gaia} to our model white dwarfs before making the same selection as presented in equation \ref{equation:Gaia}.  Using the proper motion we calculate the reduced proper motion using equation \ref{equation:RPM} and select those stars whose tangential velocity is greater than the 20\,kms$^{-1}$ curve (see Figure \ref{fig:RPMD}).

\begin{figure*}[!t]
  	
  	\includegraphics[angle=0,width=\textwidth]{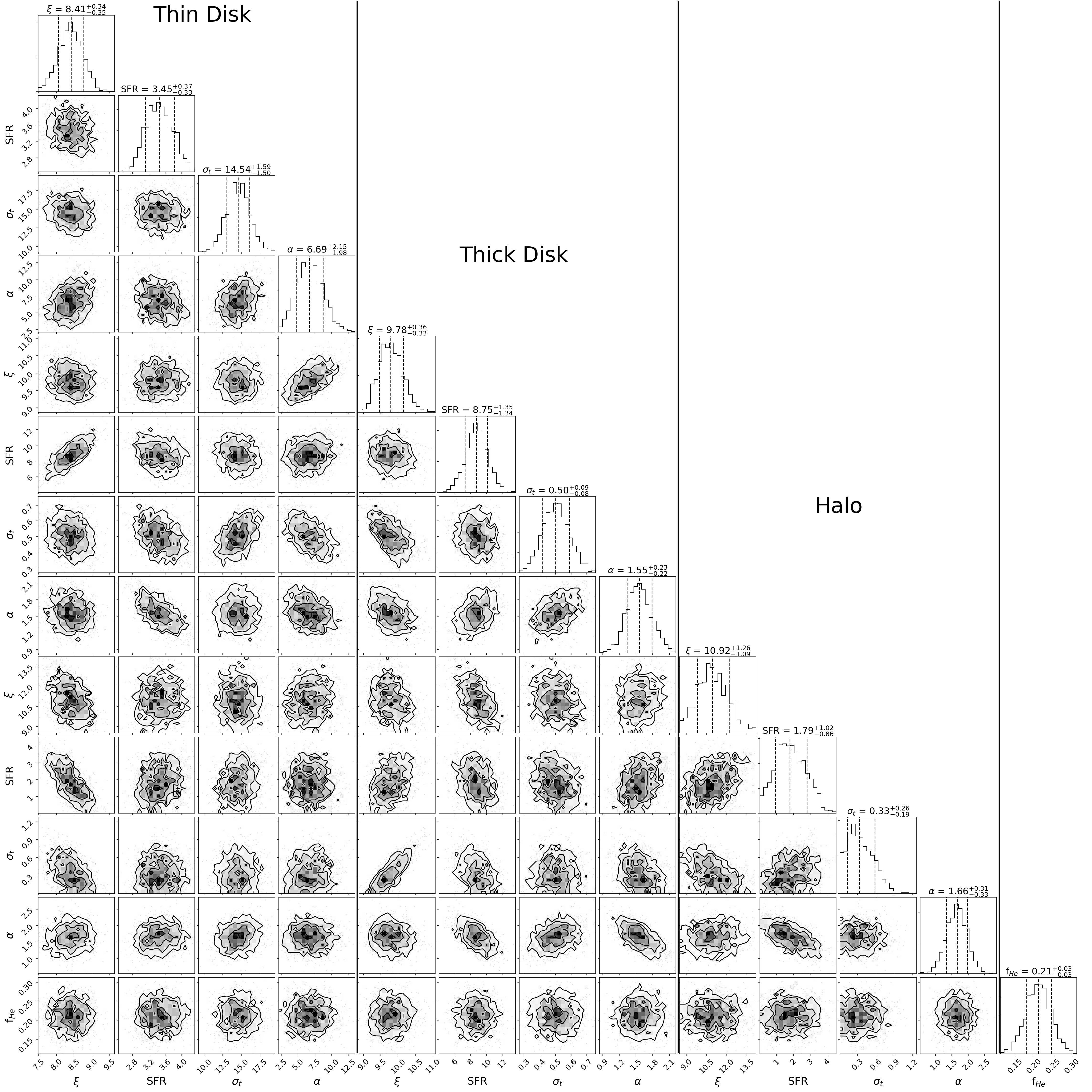}
  	\caption{Posterior distributions for the 13-dimensional parameter space sampled with \textit{astroABC}. From left to right we show the mean functional age ($\xi$, Gyr), star formation rate (SFR, M$_{\odot}$yr$^{-1}$), standard deviation ($\sigma_t$), and skew ($\alpha$) for the thin disk, thick disk, and halo. See equation \ref{equation:skew} for the definition of each parameter. The final histogram shows the fraction of white dwarfs with helium atmospheres, $f_{\textrm{He}}$. \bigskip}
  	\label{fig:corner}
 \end{figure*}
 
 \begin{figure*}[!t]
	
	\includegraphics[angle=0,width=.99\textwidth]{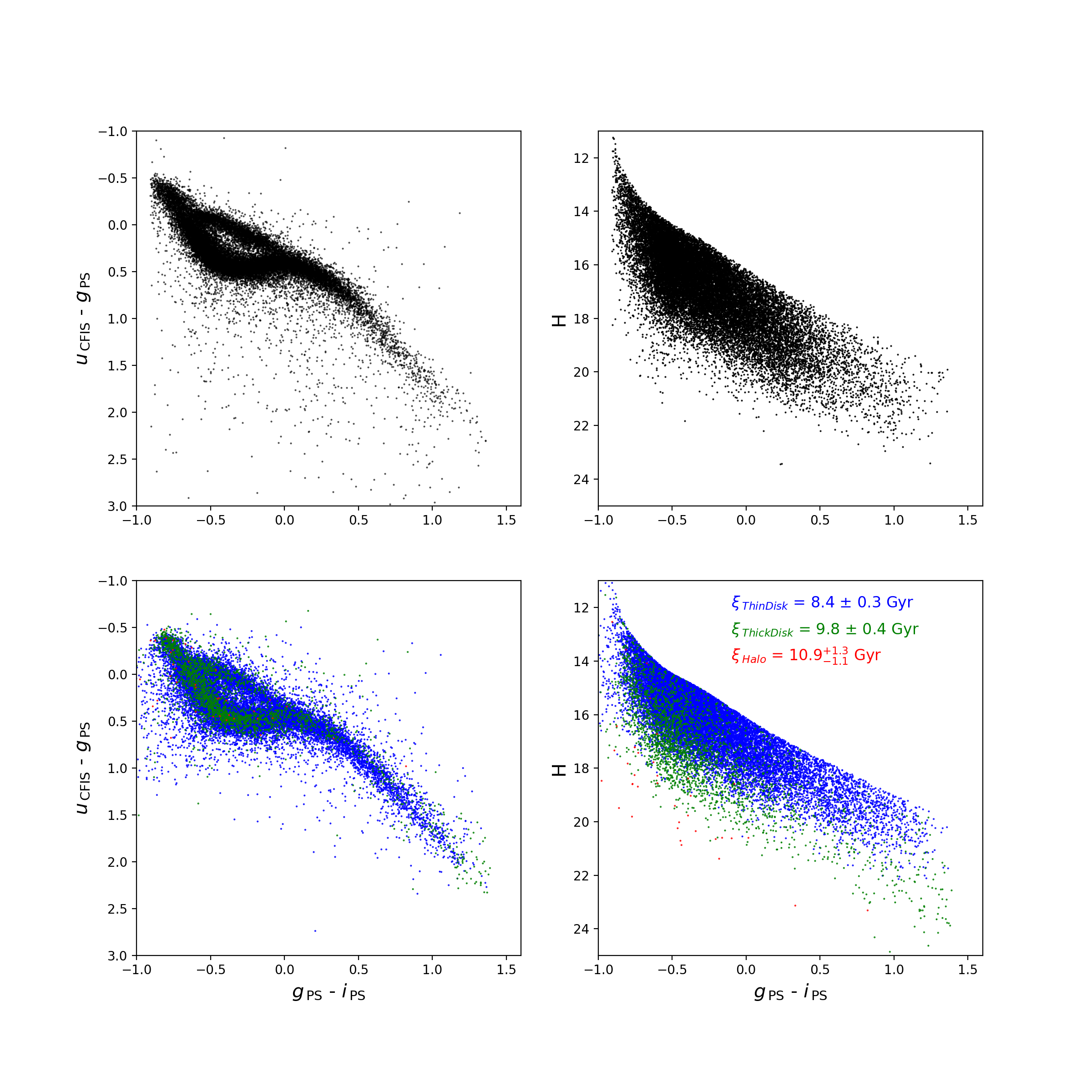}
	\caption{Comparing our best fit model (bottom panels) to the data (top panels) in both color-color (left column) and RPMD (right column). The best fit ages are shows in the bottom right panel.
		\bigskip }
	\label{fig:result}
\end{figure*}
 
We use \textit{astroABC} \citep{astroabc}, a python based approximate Bayesian computation (ABC) Markov chain Monte Carlo (MCMC) approach in order to calibrate our model. This so-called `likelihood free' method is useful in cases where a likelihood function is difficult to calculate, or unknown altogether. Unlike traditional MCMC methods used in astronomy \citep[see, e.g. \textit{emcee};][]{Emcee}, ABC MCMC does not calculate an explicit likelihood but instead relies on forward modeling. Given a set of model parameters, a simulated dataset is produced and compared to the data in order to either accept or reject the proposed parameters. More specifically, if we draw parameters $\theta$, we accept these parameters if $\rho(D - D^{*}(\theta)) < \epsilon$, where $\rho$ is the distance metric between the dataset, $D$, and the simulated data given $\theta$, $D^{*}$($\theta$), and $\epsilon$ is called the tolerance threshold. The distance metric used to compare the data and the model is
\begin{equation}
\label{eq:distance_metric}
\rho(D_i - D^{*}_i(\theta)) = \sum_{i} \frac{(D_{i} - D^{*}_i(\theta))^2}{2\sigma_{i}^{2}}.
\end{equation} 
The result at each step is a set of independent samples below a generated tolerance threshold \citep{Marjoram2003}. 

\begin{figure*}[!t]
	
	\includegraphics[angle=0,width=.99\textwidth]{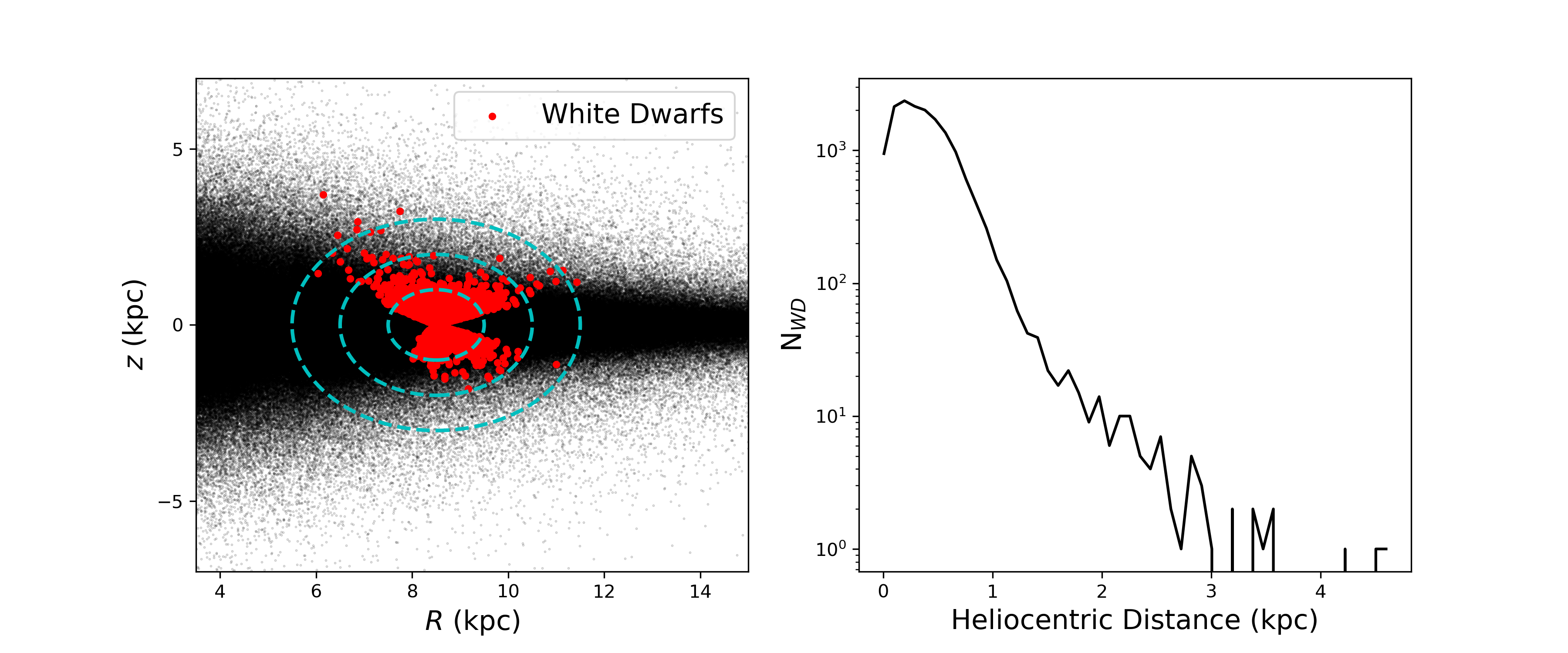}
	\caption{\textit{Left:} Positions of our model white dwarfs (red) within the Milky Way (black). The concentric cyan rings represent heliocentric distances of 1, 2, and 3\,kpc respectively. \textit{Right:} Heliocentric distance distributions of our model white dwarfs. Our sample consists mainly of local white dwarfs, with 96\% having a distance less than 1\,kpc. 
		\bigskip }
	\label{fig:MWarea}
\end{figure*}

The algorithm starts off with a large threshold and shrinks it at each step in order to approximate the PDF. When this threshold is exactly zero, then one is sampling directly from the posterior PDF; however, in that case, the acceptance fraction declines sharply as it is difficult to perfectly simulate a dataset. On the other hand, if $\epsilon$ is too large, then the algorithm is sampling from the prior. Thus, the choice of $\epsilon$ determines how much of an approximation the resulting PDFs are while minimizing the required total computational time  \citep{Marjoram2003}.

As described in \cite{Marjoram2003}, ABC methods typically require lower-dimensional summary statistics, $S(D)$, which are representative of the whole dataset in order to reduce computational costs. This summary statistic must encompass any required information contained within the data itself without neglecting anything important. Similar to \cite{Mor2018}, we use a combination of binned colors, magnitudes, and proper motions. Specifically, we compare our real and simulated data using a 3-D histogram composed of a color-color diagram (\textit{g}-\textit{i}, \textit{u}-\textit{g}), and the reduced proper motion, \textit{H}. This combination of colors and proper motions was chosen in order to encompass both age and SFR information (colors), as well as the contributions from each component (apparent magnitudes and proper motions via \textit{H}). Furthermore, the color-color diagram contains information regarding the He fraction in our sample, since they are well separated due to the exquisite \textit{u}-band photometry provided by CFIS (see Figure \ref{fig:RPMD}) Thus, in equation \ref{eq:distance_metric}, \textit{i} represents each bin in the histogram, and $\sigma$ is the sum in quadrature of the uncertainty in both the model and data in each bin. For instances where the model predicts a star, but we do not observe one, we use substitute $D_{i}$+1 into equation \ref{eq:distance_metric} in order to penalize the model by increasing the distance. We then run \textit{astroABC} with 1000 particles to obtain the approximate PDFs for the age, duration, and SFR for the three Galactic components, as well as the He fraction. We use uniform priors for every variable, with the only constraint being that the age of the components cannot exceed the age of the Universe, which we set to 13.8\,Gyr \citep{Planck2018}, and that the scale parameters, $\sigma_t$, must be greater than zero.

\section{Results and comparison to the literature}
\label{sec:Results}

The resulting corner plot containing the PDFs can be seen in Figure \ref{fig:corner}. The results from our fit can be seen in the bottom panels of Figure \ref{fig:result}, which show a model instance with the mean parameters as estimated by the final step of the Markov chain (the central line in the histograms from Figure \ref{fig:corner}). The top two panels show the actual data for comparison. We are able to reproduce the thin and thick disk population quite well, although the halo parameters are much less constrained. This is a result of the \textit{Gaia} limiting magnitude since the intrinsic faintness and scarcity of halo white dwarfs results in very few appearing in the dataset. For reference, only 44 objects lie below the 200 kms$^{-1}$ model track in Figure \ref{fig:RPMD}, which is typically associated with the Galactic halo. This sample is smaller than that obtained with deeper SDSS photometry by \citep{Munn2017}, which was used by \cite{Kilic2017} to obtain a halo age of 12.5$^{+1.4}_{-3.4}$\,Gyr. A larger sample will be required to better constrain the halo parameters --- in particular, a sample that includes objects beyond the turnover in the halo luminosity function. This will require deeper proper motion surveys, such as PS1 DR2, which will be investigated in a future paper. Within the next decade, upcoming imaging surveys such as \textit{WFIRST}  \citep{WFIRST} and LSST \citep{lsst} will provide proper motions to unprecedented depths, allowing for a more comprehensive study of the Galactic halo.

\subsection{Our Sample, in the Context of the Milky Way}

In order to properly compare with previous results, we begin by showing the volume that is sampled by our model. The resulting sight-lines in both \textit{R} (the distance from the center of the MW) and \textit{z} (the distance from the mid-plane) can be seen in the left-hand panel of Figure \ref{fig:MWarea}. The cyan dashed lines show heliocentric distances of 1, 2, and 3\,kpc, respectively, showing that the vast majority of our sample lies within one kpc of the Sun. This is confirmed in the right-hand panel which shows the distance distributions of our model white dwarfs. The result is that only 4\% of white dwarfs in our sample lie at distances greater than 1\,kpc, with a median distance of 388\,pc. Given these values, we consider our sample to be representative of the solar neighborhood, and not necessarily the whole disk.

\subsection{Star Formation History}

\begin{figure*}[!t]
	
	\includegraphics[angle=0,width=.49\textwidth]{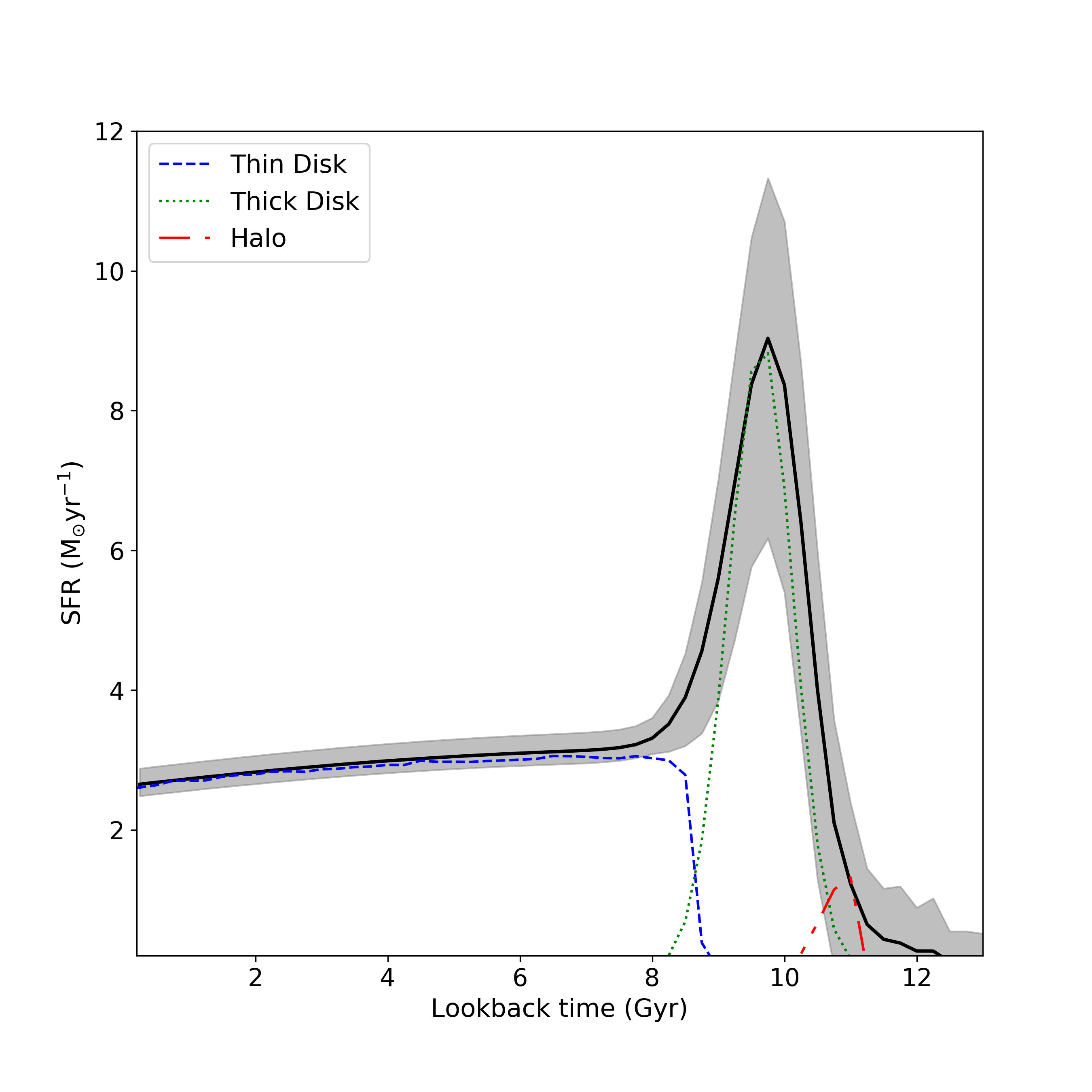}
	\includegraphics[angle=0,width=.49\textwidth]{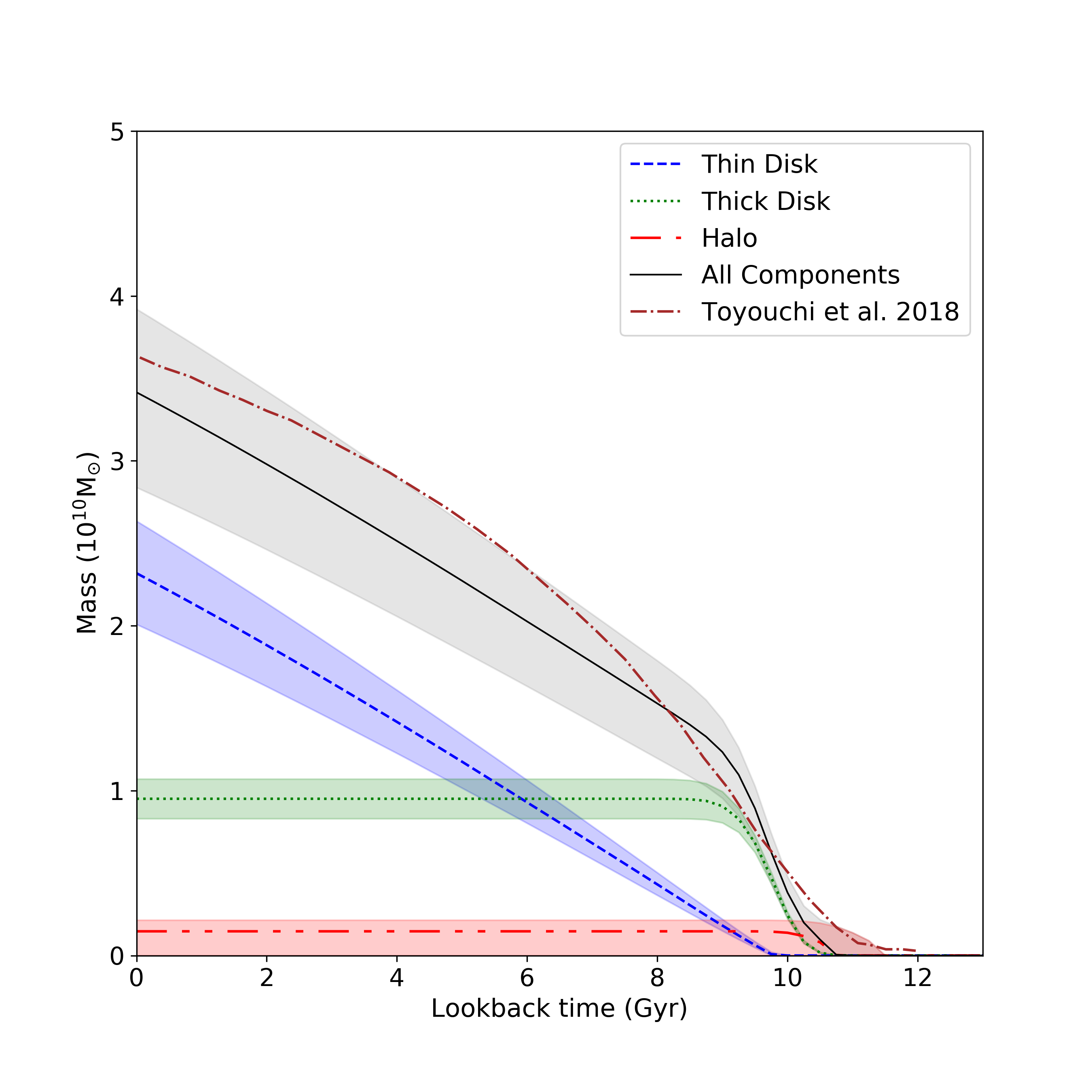}
	\caption{\textit{Left}: Milky Way star formation rate as a function of lookback time. \textit{Right}: The cumulative mass as a function of lookback time with the contribution from the thin disk (dashed blue), thick disk (dotted green) and halo (double dot-dash red) highlighted. 
		\bigskip }
	\label{fig:SFH}
\end{figure*}

\begin{figure*}[!t]
	
	\includegraphics[angle=0,width=.5\textwidth]{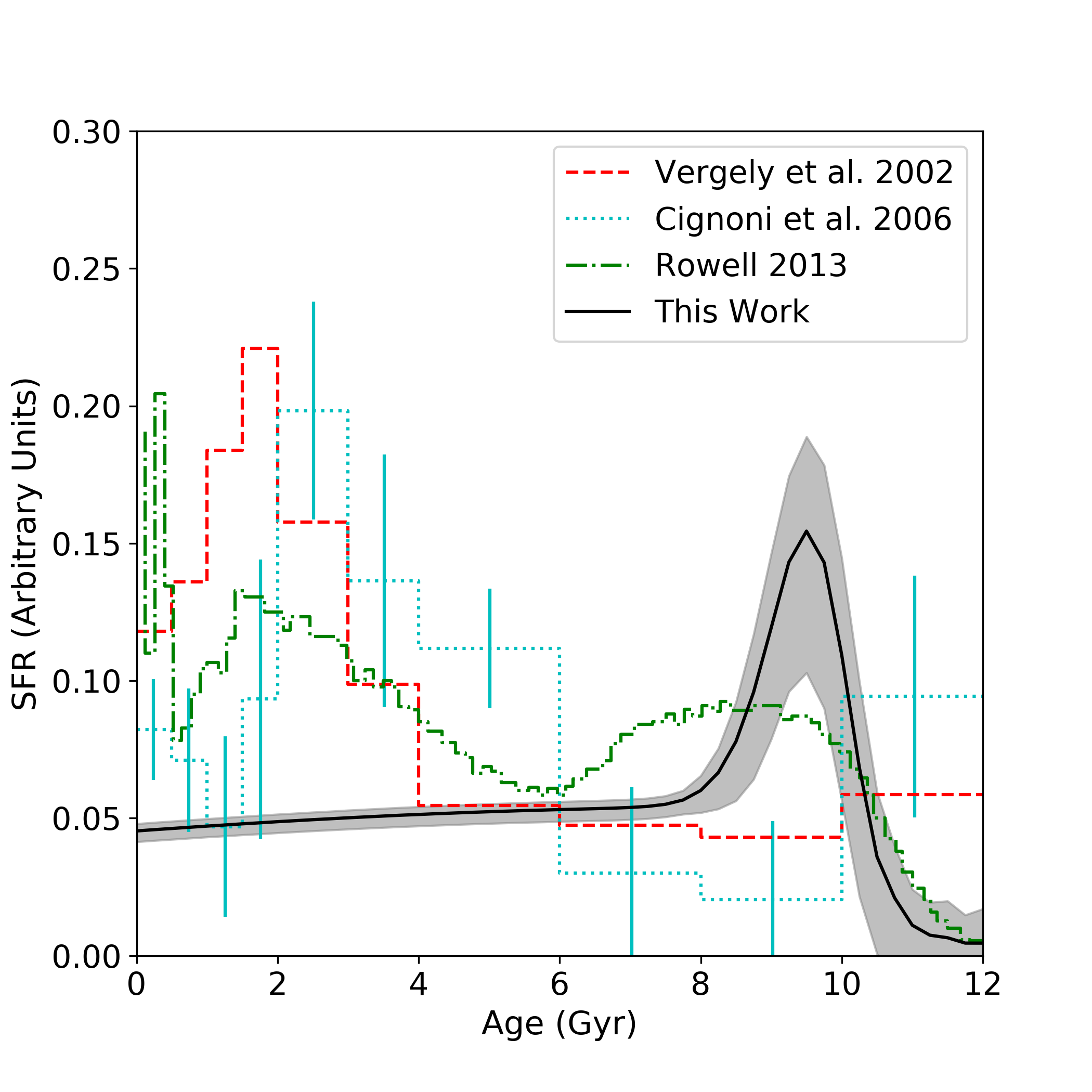}
	\includegraphics[angle=0,width=.5\textwidth]{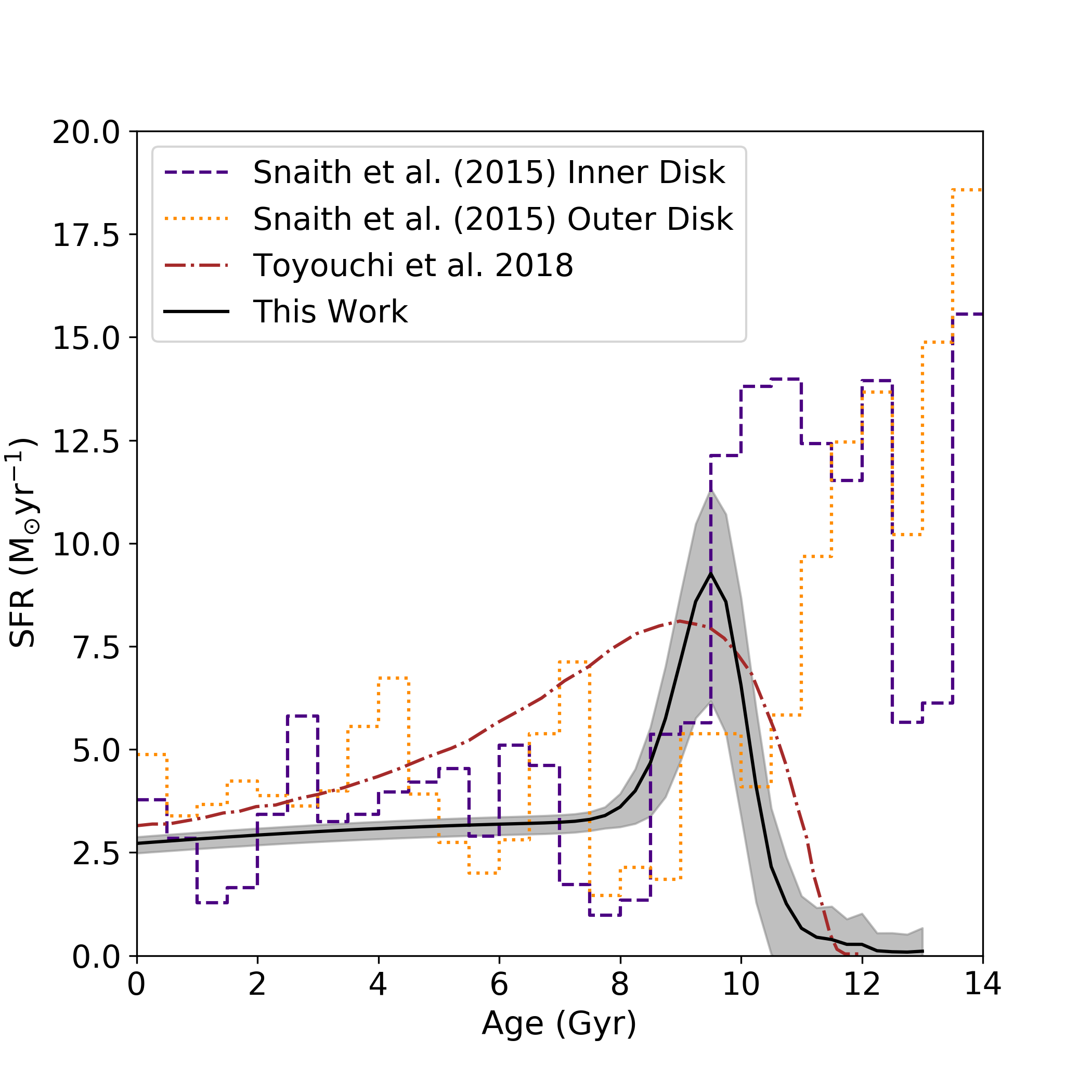}
	\caption{\textit{Left:} Our star formation history (black) compared to $Hipparcos$ results by \cite{Vergely2002} (dashed-red) and \cite{Cignoni2006} (dotted-cyan). Also shown is the result of \cite{Rowell2013}, who inverted the white dwarf luminosity function of \cite{Harris2006} (green).  \textit{Right:} Comparing our star formation history to results obtained via Galactic chemical evolution models from \cite{Snaith2015} and \cite{Toyouchi2018}. Due to the varying units presented by each study, the SFHs have been normalized, and therefore only the shape should be compared.
		\bigskip }
	\label{fig:comparison}
\end{figure*}

Our resulting SFH, integrated over the entirety of each component, can be seen in the left-hand panel of Figure \ref{fig:SFH}, where we have sampled the resulting PDFs and plotted the mean (black) and 1$\sigma$ contours (gray). Our results show an initial period of thick disk star formation, lasting 3\,Gyr and peaking at (9.8 $\pm$ 0.3)\,Gyr. This is followed by a nearly 1\,Gyr decline in star formation before finishing with a nearly constant SFR for the thin disk at approximately 3.5\,M$_{\odot}$yr$^{-1}$. Below, we present a comparison between our SFH and those found in the literature. We make note of a caveat to these comparisons, which is that we assume that each component is defined by a unimodal SFH. Thus, if a given component contains multiple bursts, we will fit the average through the bursts (see Section \ref{sec:Discussion} for further discussion).

The star formation and, in turn, the mass assembly, history of the MW has been an open question for decades. With the increase in accurate distances in the Solar neighborhood from surveys like $Hipparcos$ \citep{Hipparcos}, the local SFH could be derived by comparing the synthetic color-magnitude diagrams (CMD) to observations. This process was performed by \cite{Vergely2002} using $Hipparcos$ stars brighter than $V$ = 8, as well as \cite{Cignoni2006} who used all $Hipparcos$ stars within 80\,pc. Both of their samples are dominated by thin disk objects given their selection criteria, and in turn, they show a slowly rising star formation until it hits a peak 2\,Gyr ago. These SFHs can be seen in the left-hand panel of Figure \ref{fig:comparison} as the red dashed line and dotted cyan line respectively. In this figure, we have scaled our star formation history and hence only the shapes should be compared. We note that neither SFH recovers a strong peak early-on in the formation of the MW (the \cite{Cignoni2006} result has a slight uptick in the 10$-$12\,Gyr bin) and this may be a result of their samples being dominated by younger thin disk objects.

 We also compare our SFH to that obtained by \cite{Rowell2013} who implemented an algorithm to invert the WDLF in order to derive the local star formation history. The resulting SFH, computed from the \cite{Harris2006} WDLF, is shown as the dot-dashed green line in the left-hand panel of Figure \ref{fig:comparison}. Both SFHs are normalized, and given the varying units obtained by each study, only a comparison of the shape is appropriate. Both SFHs show two stages of star formation, however, we do not see a recent peak 2$-$3\,Gyr ago, and the peak in our SFR attributed to the thick disk happens about 2\,Gyr earlier. This is likely due to the larger volume sampled by our CFIS-PS1-\textit{Gaia}  sample, which in turn will allow us to better constrain this time-frame since our sample will contain a larger fraction of thick disk objects. It is worth noting that similarly a burst in the SFR at 600 Myr is seen in the luminosity function of the local 40\,pc white dwarf sample from \cite{Torres2016}, but not in the deeper luminosity function generated by \cite{Munn2017}. \cite{Kilic2017} concluded that this discrepancy is a result of the local sample being confined to the mid-plane of the MW, where most recent star formation occurs.

In the right-hand panel of Figure \ref{fig:comparison}, we compare our results to other SFHs found in the literature through GCE models. \cite{Toyouchi2018} used an open-box model, allowing for the inflow and outflow of gas, in order to reproduce the radial metallicity distribution obtained using APOGEE data in \cite{Hayden2015}. This result was obtained using a sample of red-giant stars with  3 $< R <$ 15\,kpc and $|z|$ $<$ 2\,kpc, which is a much larger volume than our sample. Their result shows a rise in star formation beginning at 12\,Gyr with a peak at 9\,Gyr. This is followed by a roughly constant SFR to the present day. Our SFH matches the \cite{Toyouchi2018} results quite well at recent and ancient times but differs slightly at intermediate ages.

Also plotted are results by \cite{Snaith2015} who compared their GCE model to a sample of abundances acquired as part of the HARPS survey. They model the MW disk with two components: an inner ($R <$ 7\,kpc) closed-box and an outer ($R >$ 9\,kpc) component which can accrete gas. Their resulting inner disk SFH shows a two-phase formation scenario separated by a period of inactivity lasting roughly one Gyr. A follow-up paper by \cite{Haywood2018} showed that the inner disk model is compatible with APOGEE data, and argued that the gap could be caused by the formation of a bar. 

Their inner and outer disk results have been plotted in purple and orange, respectively, in Figure \ref{fig:comparison}. Given the relative nature of their presented star formation history, we have scaled their SFH so that both periods associated with the thin disk are equal. The inner disk shows an initial burst of star formation, which they attribute to the thick disk, and ends with a relatively constant SFR over the past 7\,Gyr associated with the thin disk. Comparing the general shape of their inner disk SFH, an initial peak of SFH followed by a decline to a nearly constant SFR, shows that the two models are consistent, however, we do not \textit{require} a gap in our SFH. This is likely because this is a feature of the inner disk, and since the gap is thought to be a result of the formation of a bar, it did not have a large effect on the local star formation we measure.

Our results differ in terms of the SFH of the thick disk, particularly at the earliest stages of the MW's formation. This is likely a consequence of the large uncertainties in both models at the earliest epochs of the MW and in particular the difficulty of detecting white dwarfs with cooling ages exceeding 11\,Gyr. This could also be a result of our choice of scale height (See section \ref{sec:Discussion}), as an increased scale height at these epochs would increase our SFR.

\cite{Haywood2019} argue that the solar neighborhood is better described by the outer disk evolution, given the enrichment history of the outer disk. In their outer disk scenario, their simulation includes a single accretion event at 10 Gyr (and thus the SFR between 10 and 13.8\,Gyr is unconstrained) and results in a consistent SFR over the past 10\, Gyr, similar to the inner disk. This is roughly consistent with our results, particularly within the past 8\,Gyr.

\subsection{Component Masses}

The right-hand panel of Figure \ref{fig:SFH} shows the cumulative mass of the MW as a function of time, again with all 1000 particles shown in gray. This is done by integrating our SFH and density functions over the entire MW. We measure a total stellar mass of (3.4 $\pm$ 0.6) $\times$ 10$^{10}$\,M$_{\odot}$ for our thin disk, thick disk, and stellar halo. Also plotted is the result of \cite{Toyouchi2018} which is consistent within uncertainties with our result. Given the relative nature of the other star formation histories, we have chosen not to make a direct comparison with the results plotted in Figure \ref{fig:comparison}. 

The mass of the MW's stellar disk is an important astronomical property, and as such, has been extensively studied. Our result is systematically lower than many previous results. Recent mass models of the MW have found a total disk mass between 3.6-5.5 $\times$ 10$^{10}$ M$_{\odot}$  \citep[see, e.g,][]{Gerhard2002, Flynn2006, McMillan2011,Bovy2013}. It should be noted, however, that these estimates are highly dependent on the model assumptions, including the scale length and scale height, IMF, and whether or not a bulge/bar is included. Furthermore, it should be reiterated that our survey volume is rather small, and thus extrapolating our volume to the entire disk will inevitably contain a systematic uncertainty.

\subsection{Local Stellar Density and White Dwarf Number Density}

Given the systematic uncertainties associated with extrapolating our result to the entire disk, a more appropriate comparison is likely with the local stellar density. The stellar density as calculated by our model is (0.036 $\pm$ 0.004) M$_{\odot}$pc$^{-3}$ for the thin disk, (4.5 $\pm$ 0.5) $\times$ 10$^{-3}$ M$_{\odot}$pc$^{-3}$ for the thick disk, and (3.2 $\pm$ 3.1) $\times$ 10$^{-6}$ M$_{\odot}$pc$^{-3}$ for the halo. Our values agree with \cite{Bovy2017}, who used \textit{Gaia}  DR1 to calculate a main-sequence stellar density of 0.040 $\pm$ 0.002 M$_{\odot}$pc$^{-3}$.

The resulting white dwarf densities are (4.8 $\pm$ 0.4) $\times$ 10$^{-3}$\,pc$^{-3}$ for the thin disk, (1.0 $\pm$ 0.2) $\times$ 10$^{-3}$\,pc$^{-3}$ for the thick disk, and (6.3 $\pm$ 2.4) $\times$ 10$^{-6}$\,\,pc$^{-3}$ for the halo. This corresponds to a total white dwarf number density of (5.8 $\pm$ 0.5) $\times$ 10$^{-3}$\,pc$^{-3}$. Our values are consistent with the estimates of (5.5 $\pm$ 0.1) $\times$ 10$^{-3}$\,pc$^{-3}$ from \cite{Munn2017}, and marginally higher than results of 4.6 $\times$ 10$^{-3}$\,pc$^{-3}$ from \cite{Harris2006}, (4.8 $\pm$ 0.4) $\times$ 10$^{-3}$\,pc$^{-3}$ from \cite{Torres2019}, and (2.81 $\pm$ 0.52) $\times$ 10$^{-3}$\,pc$^{-3}$ from \cite{Fantin2017}.


Our resulting white dwarf contributions within the solar neighborhood break down to (83 $\pm$ 5)\,\%, (17 $\pm$ 3)\,\%, and (0.11 $\pm$ 0.05)\,\% from the thin disk, thick disk, and stellar halo respectively. \cite{RowellHambly2011} constructed thin disk, thick disk, and halo luminosity functions using a statistical approach based on kinematics and photometry from the SuperCOSMOS survey in order to obtain local fractions of 79, 16, and 5 percent, respectively. This result was consistent with \cite{Reid2005} who found that thick disk white dwarfs should comprise roughly 20\% of the local population. Our thick disk fraction is marginally lower than the recent result by \cite{Torres2019} who obtained fractional contributions of 74, 25, 1 \% from the thin disk, thick disk, and halo respectively.

\subsection{Component Ages}

The white dwarf luminosity function has been used as a tool to calculate the ages of various populations within the MW, including its field population. Our functional ages, $\xi$ (see Equation \ref{equation:skew}), for the thin disk and thick disk are 8.41$^{+0.34}_{-0.35}$\,Gyr and 9.78$^{+0.36}_{-0.33}$\,Gyr, respectively. Since the age can be defined as the onset of star formation, we also present these values as (8.5 $\pm$ 0.3)\,Gyr for the thin disk and (11.3 $\pm$ 0.3)\,Gyr for the thick disk. 

The first age measurement using white dwarfs was preformed by \cite{Winget1987} who obtained an age of (9.3 $\pm$ 2.0\,Gyr for the Galactic disk. Following this, \cite{Leggett1998} used spectroscopic data from a sample of 43 white dwarfs to obtain an age of (8 $\pm$ 1.5)\,Gyr for the disk. Both of these samples were likely dominated by thin disk objects, and this is reflected in their age measurements.

Our SFH is consistent with those derived from the oldest open clusters, which are typically associated with the thin disk. \cite{Bedin2005} and \cite{Garcia2010} used the WDLF to determine an age of $\sim$8\,Gyr for NGC 6791, an open cluster with solar-like metallicity. This cluster likely formed at the onset of thin disk star formation given its age and metallicity. 

Our thick disk age is consistent with metal-rich globular clusters, like 47 Tucanae, which have metallicities comparable to thick disk field stars ([Fe/H] = $-$0.75). \cite{Hansen2013} used \textit{Hubble Space Telescope} photometry to observe the WDLF and obtain an age of (9.9 $\pm$ 0.7)\,Gyr. In our star formation history, this would place its formation near the peak of the thick disk star formation period.

\cite{Kilic2017} used the WDLF presented by \cite{Munn2017}, who performed second epoch follow-up photometry within the SDSS footprint to calculate proper motions and select their white dwarf sample. The ages of our thin disk and thick disk are consistent within the errors with their results of 7.4$-$8.2\,Gyr and 9.5$-$9.9\,Gyr. 

Although less constrained than our disk ages, our resulting functional halo age is 10.92$^{+1.26}_{-1.09}$\,Gyr. This translates into an onset age of (12.3 $\pm$ 1.3)\,Gyr. This result is consistent with metal-poor globular clusters associated with the stellar halo, which typically have ages of 11$-$13\,Gyr. For example, \cite{Hansen2013} obtained an age of (11.7 $\pm$ 0.3)\,Gyr for NGC 6397 and \cite{Bedin2009} found an age of (11.6 $\pm$ 0.6)\,Gyr for M4, both using the WDLF. Our result is also consistent with \cite{Kilic2017}, who presented an age of 12.5$^{+1.4}_{-3.4}$\,Gyr. 

\begin{figure*}[!t]
	
	\includegraphics[angle=0, width=\textwidth]{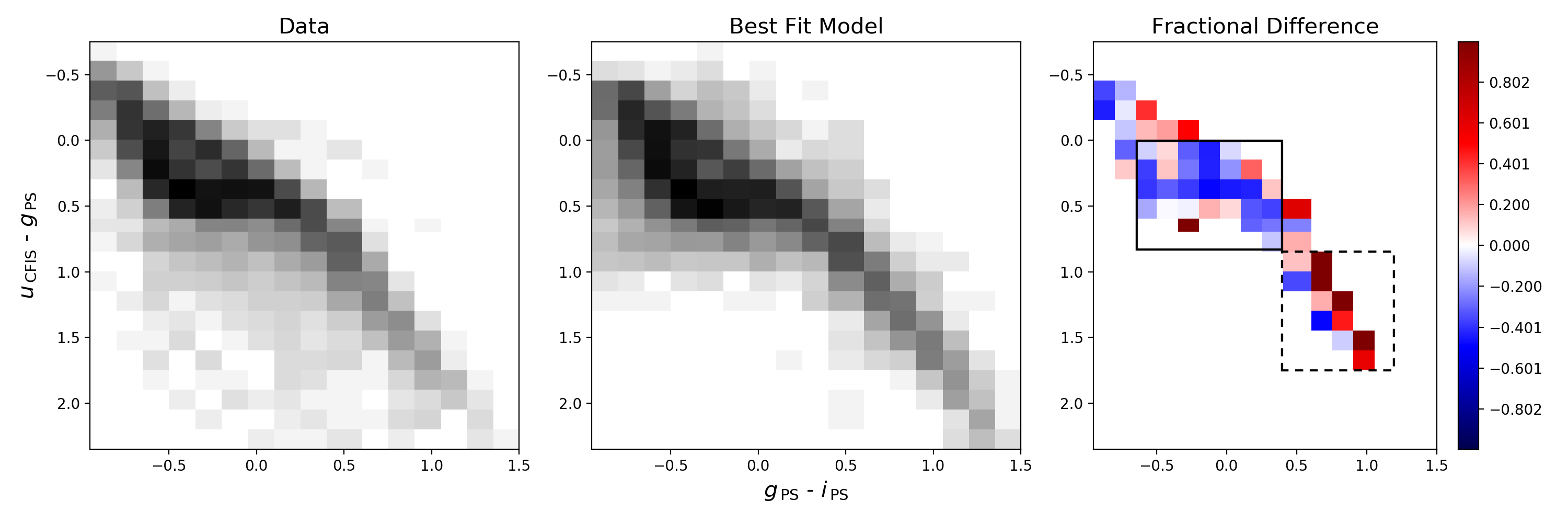}
	\caption{Color-color diagram showing the data (left) and best-fit model (center) binned every 0.15 mags. The right-hand panel shows the difference between the data and the model, color-coded by either an excess (blue) or deficit (red) within the model. The black box shows the main location where the model contains a deficit of white dwarfs relative to the data, and these objects have a mean formation age of (3.3 $\pm$ 1.8)\,Gyr. The dashed box represents the location where our model over-predicts the number of white dwarfs, and these objects have a mean formation age of (5.8 $\pm$ 1.1)\,Gyr. This suggests a more bimodal formation history, with a 50\% increase in SFR near 3\,Gyr and a 30\% deficit at 6\,Gyr. }
	\label{fig:residual}
\end{figure*}

\subsection{He Fraction}

The resulting He fraction in our sample is (21 $\pm$ 3)\%. This is consistent with \cite{Bergeron2011}, who fit model atmospheres to a spectroscopic white dwarf sample from the Palomar Green Survey. They showed that approximately 20\% of white dwarfs with temperatures below 20,000 K appeared to contain a helium atmosphere. This result was followed up by \cite{Genest2019}, who used SDSS spectra to show that the fraction of DB stars increases at lower temperatures (see their Figure 23). This fraction was found to be approximately 5\% at temperatures greater than 20,000 K, with an increase to 20$-$25\% at 12,000 K. Using a sample of local white dwarfs \cite{Limoges2015} also showed that 25\% of their sample was best described by models containing a helium atmosphere. 

Our measurement is also consistent with recent results from \cite{Kilic2018} who used a color-magnitude diagram, generated using SDSS photometry and \textit{Gaia}  DR2 astrometry, to obtain a DB fraction of 36 $\pm$ 2\% for a sample of white dwarfs within 100\,pc. Since He-atmosphere white dwarfs are most pronounced at higher temperatures their result was quoted within a region of color-magnitude space encompassing the bifurcation of DA and DB white dwarfs due to the increased strength of the Balmer lines. In the temperature region, He-atmosphere white dwarfs are classified as type DB since He I lines are present in their spectra. Applying the same selection area, $M_{u_{\textrm{SDSS}}}$ between 10 and 14 and \textit{u$_{\textrm{SDSS}}$} - \textit{g$_{\textrm{SDSS}}$} between $-$0.4 and +0.6, to our CFIS and PS1 model color-magnitude diagram produces a consistent DB fraction of 34 $\pm$ 3\%.

\section{Discussion}
\label{sec:Discussion}

As Figure \ref{fig:SFH} shows, our results suggest the star formation of the MW disk began (11.3 $\pm$ 0.3)\,Gyr ago. Following a brief period of halo formation, the thick disk rapidly begins to form stars until it reaches a peak (9.8 $\pm$ 0.3)\,Gyr ago. This period is followed by a decline in thick disk star formation. The thin disk then begins to form stars (8.5 $\pm$ 0.4)\,Gyr ago. 

The transition between the thin and thick disk SFHs was a fairly active period in the MW's history. Recent results by \cite{Helmi2018} and \cite{Belokurov2018} provided evidence that this epoch was dominated by the accretion of at least one satellite galaxy, and may have contributed to the rapid transition between the thick and thin disk. The resulting merger would have heated up the existing stars to form a thick disk approximately 10 Gyr ago \citep{Helmi2018}. Given that a merger can enhance the SFR for a short period before quenching \citep{DiMatteo2008}, our results suggest that the merger occurred shortly before our peak SFR  at (9.8 $\pm$ 0.3)\,Gyr, and before the onset of star formation in the thin disk. Thus, this scenario is consistent with a merger triggering the epoch of peak star formation in the Milky Way and the start of star formation in what we now call the thin disk.

In this section, we discuss the impact of our assumptions on these results, including our choice of star formation prescription, and discuss the next steps for this study.

\subsection{Effect of Star Formation Prescription}

As shown in equation \ref{equation:skew}, we assume skewed Gaussian functions for the three components, as allowing a skewness increases the degrees of freedom of the function. In this sense, we have assumed unimodal functions for the star formation histories, and we will inevitably fit over any potential multimodal peaks in the star formation. Specifically, if the star formation history of a single component is composed of multiple bursts, then we will fit the average star formation throughout these periods of increased and decreased star formation. To examine this possibility, we have plotted the fractional residual between the data and the best fit model in Figure \ref{fig:residual}.

Figure \ref{fig:residual} shows a deficit of white dwarfs within the solid black box and an excess of white dwarfs within the dashed box. These areas represent mean formation ages of  (3.3 $\pm$ 1.8)\,Gyr and (5.8 $\pm$ 1.1)\,Gyr respectively. This suggests that we are underestimating the star formation rate around 3\,Gyr and overestimating at 6\,Gyr by roughly 50\% and 30\%, respectively. This is consistent with the results of \cite{Rowell2013} and \cite{Vergely2002}, as shown in the left-hand panel of Figure \ref{fig:comparison}, who see an increase in star formation between 2 and 3\,Gyr and a resulting decrease in the vicinity of 6\,Gyr. This result is also in line with the recent work by \cite{Mor2019}, who compared simulations of the Besan\c{c}on Galactic model with the colors, magnitudes, and parallaxes from \textit{Gaia}  DR2, which shows a burst of star formation between 2 and 3\,Gyr in the past. The slight bimodality suggested by Figure \ref{fig:residual} is, however, not seen in the GCE models of \cite{Snaith2014} and \cite{Toyouchi2018}, as seen in the right-hand panel of Figure \ref{fig:comparison}.

\subsection{Effect of the Scale Height}

\begin{figure*}[!t]
	
	\includegraphics[angle=0, width=\textwidth]{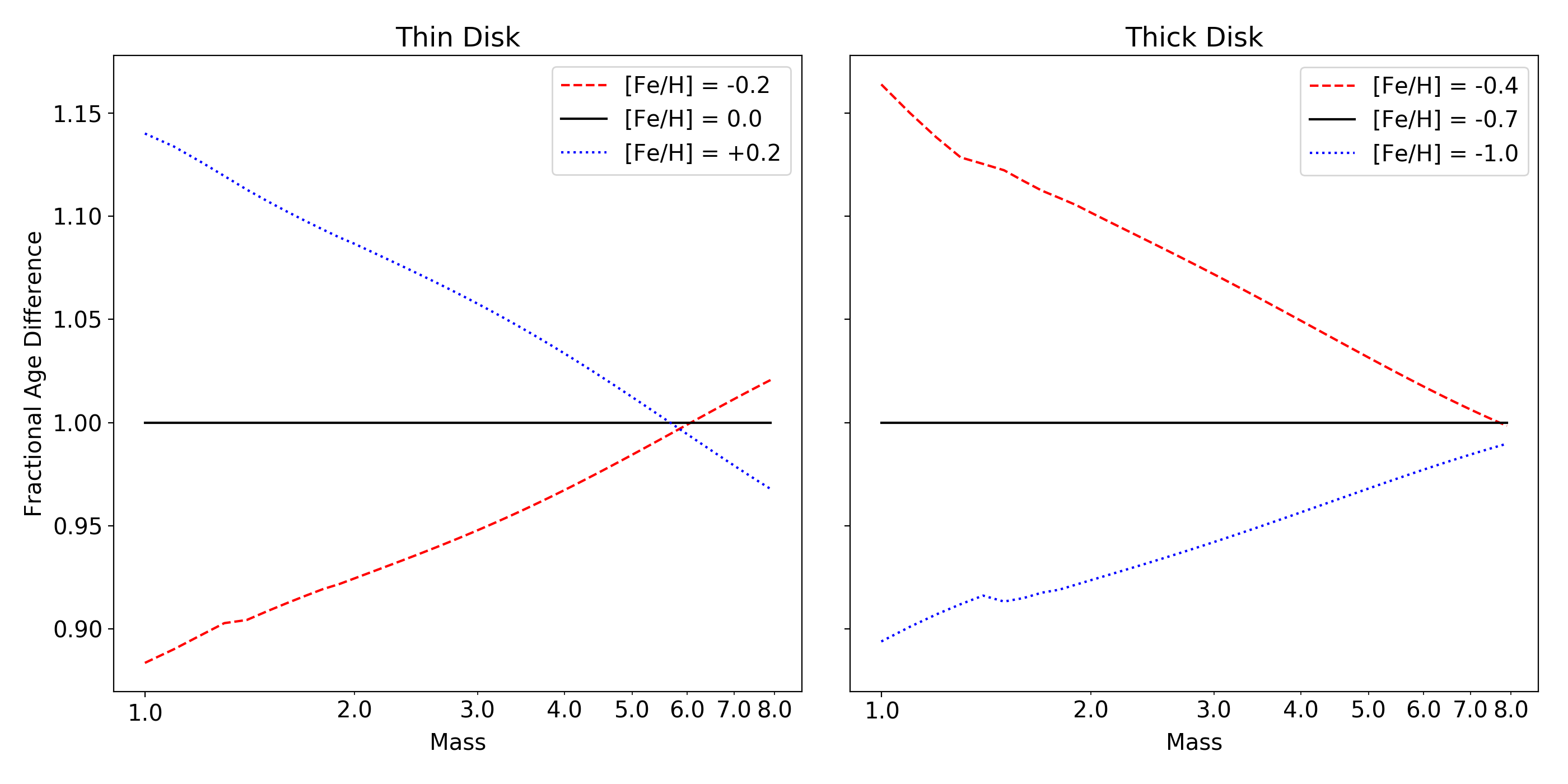}
	\caption{ Fractional difference between the mean metallicity (black solid line) for the thin disk (left) and thick disk (right) for a star with higher metallicity (dotted blue) and lower metallicity (dashed red) than the mean. }
	\label{fig:metallicity}
\end{figure*}

In our model, we have assumed a three-component Galaxy with constant scale heights. This assumption will inevitably impact our SFH, as a decrease in the scale height will result in a lower star formation rate. This is because more stars will reside closer to the plane, and will have a higher probability of passing our observational criteria. 

\cite{Robin2014} found that the scale height for the thick disk  (described using a sech$^2$ distribution) likely contracted from 800\,pc to 350\,pc between 12 and 10\,Gyr. Given this scenario, our SFH would under-predict the SFR while the scale height was greater than 550\,pc and over-predict while it was below this value. However, given the relatively small distances of the vast majority of our sample, the impact of deviating from our chosen mean of 550\,pc is minimal. Specifically, at the maximum scale height of 800\,pc we would need to produce 11\% more stars than at a scale height of 550\,pc. Similarly, at 350\,pc we would produce 17\% more stars than at 550\,pc, reducing our SFR. These values are well within our uncertainties seen in Figure \ref{fig:SFH}.

A thick disk composed of a range of scale heights has been suggested by recent results in order to explain the dip in [$\alpha$/Fe] seen in the APOGEE data \citep{Bovy2012, Haywood2013, Haywood2016}. Given that we modeled a single scale height, adding further components to our model would result in both scenarios described above. The net result would be a flattening of the thick disk SFH with a minimal decrease to the peak since the total number of white dwarfs produced would need to remain the same.

\bigskip

\subsection{Effect of Metallicity}

In our model, we have chosen to set a constant metallicity for each Galactic component. The metallicity of a given star affects the pre-white dwarf lifetime through the analytical lifetimes from \cite{Hurley2000}. Therefore, changing the metallicity of a given star will change its lifetime, and in turn its white dwarf cooling age. Specifically, for a fixed initial mass, the progenitor lifetime will increase with increasing metallicity, which in turn decreases the white dwarf cooling time. \cite{Tononi2019} found that including a dispersion in the metallicity provided a better fit to the 40\,pc white dwarf luminosity function. Adding a dispersion to our metallicity would add a dispersion to the progenitor ages, which, through the cooling age, affects the photometry. If the dispersion is symmetric the effect would be equal at all ages and thus not impact our results. We have studied this effect in Figure \ref{fig:metallicity}, which shows the fractional change in progenitor lifetime for a higher and lower metallicity relative to our chosen values for the thin and thick disk. While not perfectly symmetric, the implementation of a constant dispersion would only systematically increase our ages by a few percent, which is minimal relative to our uncertainties. 

\begin{figure}[!t]
	
	\includegraphics[angle=0, width=0.5\textwidth]{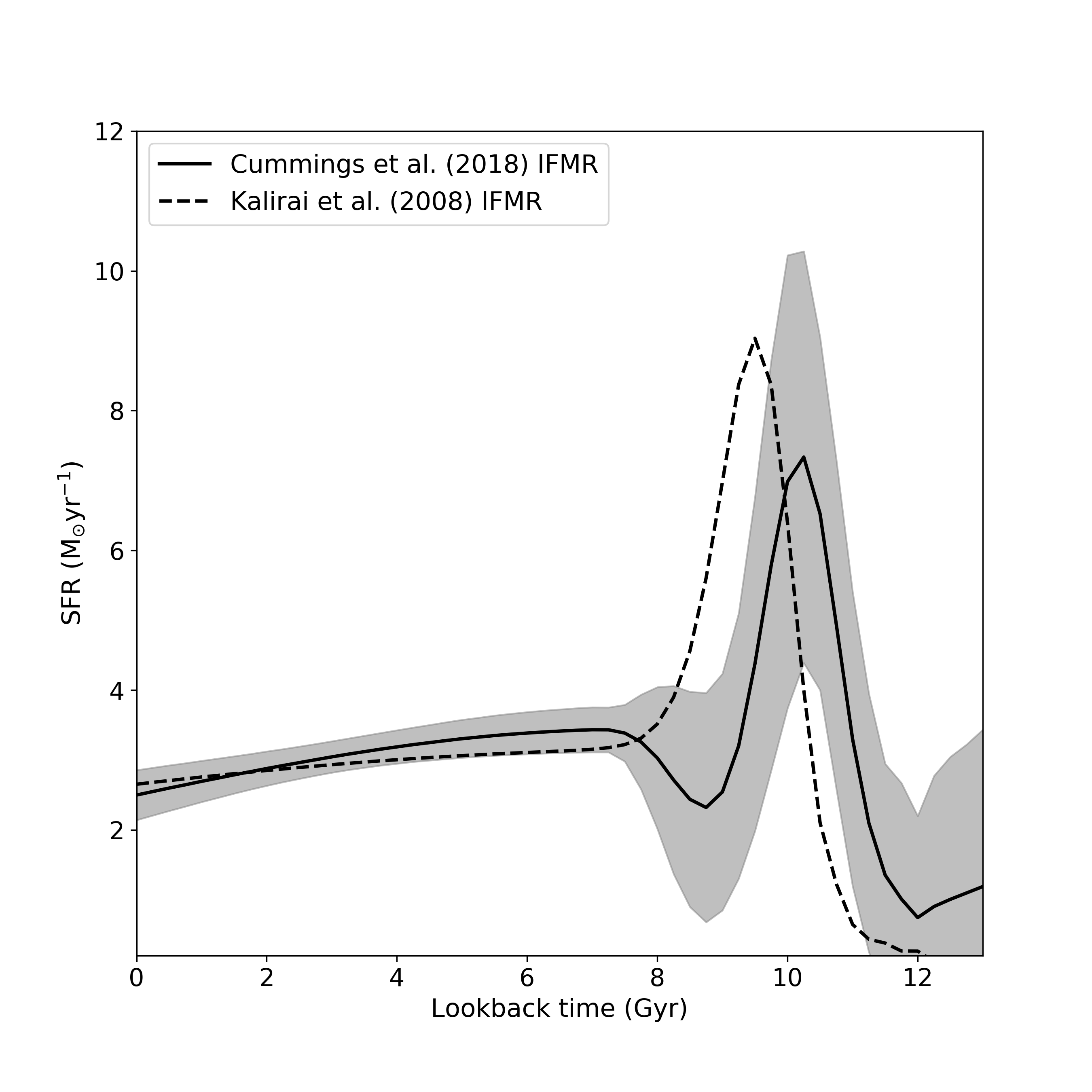}
	\caption{A comparison between the resulting star formation history using the IFMR from \cite{Kalirai2008} (dashed line) and \cite{Cummings2018} (solid line). The \cite{Cummings2018} IFMR returns systematically higher masses ($\sim$0.05\,M$_{\odot}$) which results in a 0.3\,Gyr increase in the age for the thick disk.}
	\label{fig:IFMR}
\end{figure}

\subsection{Effect of the IMF and IFMR}

Ultimately, as with any population synthesis study, our results are contingent on our assumptions. Assuming that the IMF is constant over time, changing our IMF to, for example, a Salpeter or Chabrier will affect the overall normalization, but not the shape of our SFH. This is because much of the discrepancy between IMFs is present at the low-mass end and the white dwarfs currently present in the MW were formed by stars with initial massed greater than 0.8$-$0.9 M$_{\odot}$, for which the slope of the IMF is well constrained.

\begin{figure*}[!t]
	
	\includegraphics[angle=0,width=\textwidth]{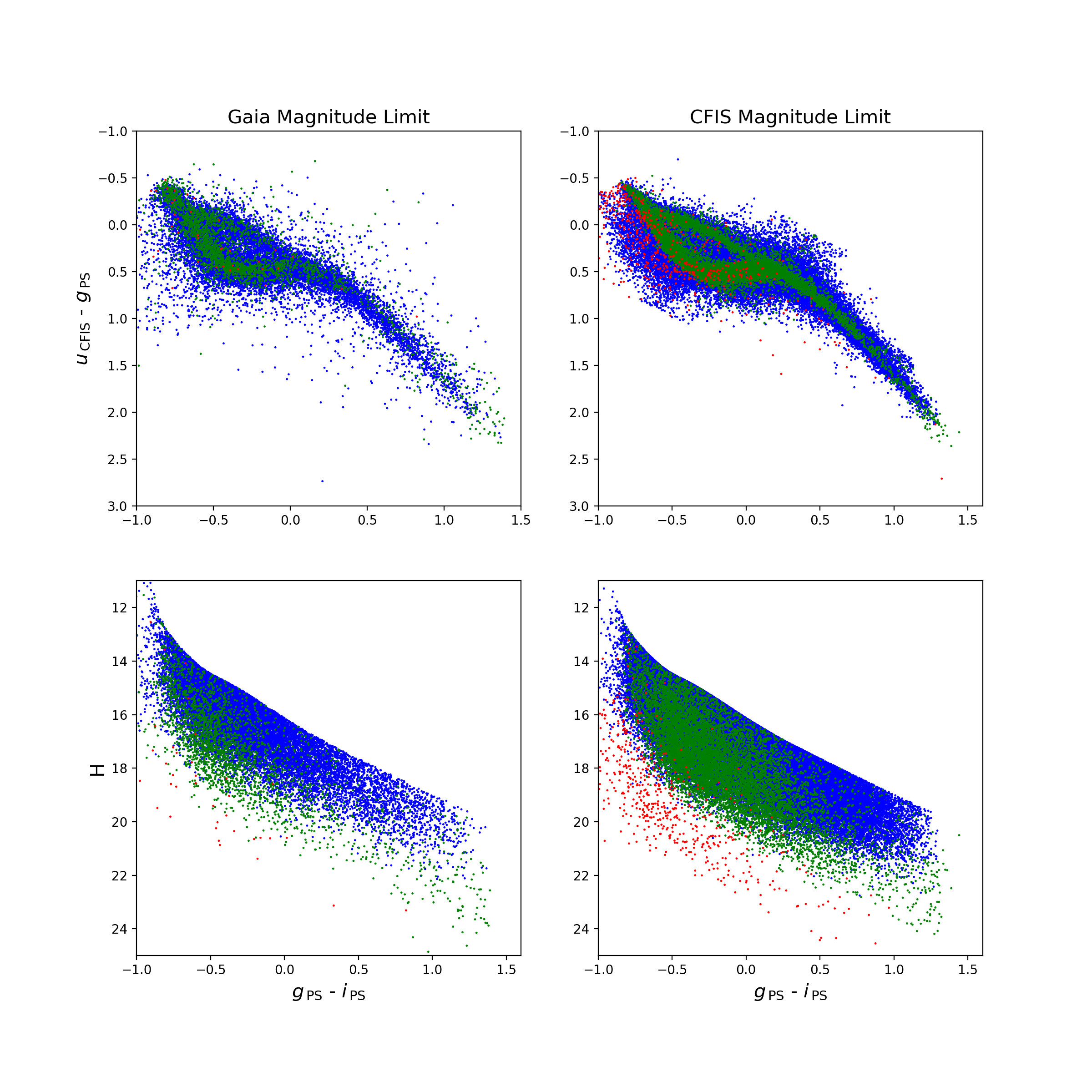}
	\caption{Comparing the resulting white dwarf populations based on the \textit{Gaia}  magnitude limit (left columns; \textit{G} = 20.7) and the CFIS magnitude limit (right columns; $u$ = 24.2). The fainter magnitude limit results in a 5$-$10$\times$ increase in halo white dwarfs relative to the CFIS-PS1-\textit{Gaia} sample.
		\bigskip }
	\label{fig:future}
\end{figure*}

Changing the IFMR will impact the ages of the white dwarfs, as the white dwarf cooling age is a strong function of the mass. As the IFMR is generally estimated from empirical data, \textit{Gaia}  DR2 can be used to calculate the IFMR. Recent work by \cite{ElBadry2018}, who used spectroscopically confirmed DAs in wide binaries, has shown promise, however extending this work to all white dwarfs has been challenging given the impact of the atmospheric type on the resulting cooling age, which requires spectroscopic follow-up. The results by \cite{ElBadry2018} and \cite{Cummings2018} also showed that much of the discrepancy between IFMRs arises at high progenitor masses, for which there are very few such stars in our observed sample (see their Figure 3). While we note that there is no consensus on an IFMR, changing the IFMR is likely to have the highest impact on the SFH at old ages where the majority of old white dwarfs are formed from higher mass progenitors.  

In order to test this, we ran a simulation with the semi-empirical IFMR derived by \cite{Cummings2018} using the MIST isochrones from \cite{Choi2016} to calculate the progenitor ages. The IFMR presented by \cite{Cummings2018} produces systematically larger white dwarf masses relative to the IFMR from \cite{Kalirai2008}, with typical differences being $\sim$0.05 M$_{\odot}$. At a fixed SFH, or equivalently, fixed white dwarf cooling time, increasing the white dwarf mass will increase the resulting temperature. The end result is that the ages must increase to match the observed colors as these white dwarfs will require more time to cool to equivalent temperatures, represented by the observed colors in our model. The resulting SFH with the \cite{Cummings2018} IFMR can be seen in Figure \ref{fig:IFMR}. The shape of the two SFHs are roughly consistent, although the resulting thick disk functional age is $\sim$0.3\,Gyr older than the result with the \cite{Kalirai2008} IFMR seen in Figure \ref{fig:SFH}.

\subsection{Effect of Atmospheric Composition}

In our model, we assumed that white dwarfs have one of two atmospheric compositions: pure-hydrogen or pure-helium. In reality, white dwarfs with helium dominant atmospheres typically contain some hydrogen \citep{Koester2015, Rolland2018}. \cite{Genest2019} showed that this additional hydrogen impacts the photometric mass determinations of helium-atmosphere white dwarfs, particularly at low temperatures where the hydrogen abundance can increase due to convective mixing. Given that our model obtains masses from the IFMR (and hence both populations have the same mass distribution), this issue will not impact our results, however, a change in the atmospheric composition will affect the rate of cooling of the helium-atmosphere white dwarfs. As discussed in \cite{Kilic2017}, pure-helium and pure-hydrogen white dwarfs take the same amount of time to cool to roughly 5,000\,K, however, hydrogen-atmosphere white dwarfs begin to cool slower than their helium counterparts below 5,000\,K. Specifically, for a 0.6 M$_{\odot}$ white dwarf with a pure-hydrogen atmosphere, it will take $\sim$1.5\,Gyr longer to reach 4,000\,K than if it had a pure-helium atmosphere. At 0.8 M$_{\odot}$, this discrepancy rises to $\sim$2.5\,Gyr. Within this temperature range, our resulting model produces very few white dwarfs, and those which pass our selection criteria have pure-hydrogen atmospheres. This is consistent with recent results suggesting that the majority of cool white dwarfs have nearly pure-hydrogen atmospheres \citep{Kowalski2006, Giammichele2012, Limoges2015, Bergeron2019}. The addition of mixed H/He atmospheres may allow a few He-dominant white dwarfs to pass the selection criteria by slowing down their cooling at this epoch, however, this effect would be minimal given the scarcity of objects in this region.

\subsection{Model Improvements}

A few straightforward improvements could be made to our white dwarf population synthesis model in the future. The first would be the inclusion of binaries within the model. Many main sequence+white dwarf binaries would not pass our selection criteria given their colors, which would increase the resulting star formation because the model would need to produce more stars to match the data.

Binary evolution during previous stages of stellar evolution can also affect the evolution of white dwarfs. Recent evidence presented by \cite{Kilic2018} suggests that there exists a population of white dwarfs that formed through the merging of two main-sequence stars, and that this population may account for as much as $15-25$\% of the field white dwarf population. This result is consistent with the theoretical predictions made by \cite{Toonen2017} using the local 20\,pc sample. Single white dwarfs formed through merging events have also been suggested by \cite{Liebert2005} to explain the high-mass bump in the DA white dwarf mass distribution, but at a rate of 12$-$15\%. This scenario will not only affect the progenitor age, but also the cooling age since higher mass white dwarfs cool more quickly than their lower-mass counterparts. This will inevitably affect the measured star formation history, however, further work will be needed to determine the exact contribution from this scenario.

Extremely low mass (ELM) white dwarfs (M $<$ 0.3\,M$_{\odot}$) are also likely present within our dataset. These white dwarfs cannot be formed through normal evolutionary channels and are thought to form as a result of extreme mass-loss before the horizontal branch, leaving an exposed core composed primarily of helium \citep[see, e.g,][]{Kilic2011, Sun2018}. \cite{Kepler2007} found that this population could account for as much as 10\% of the total white dwarf population, and they have been found in many white dwarf surveys \citep{Kleinman2013, Kepler2016, Kepler2019}. Recent work by \cite{Pelisoli2019} using \textit{Gaia}  DR2 suggests a space density of 275\,kpc$^{-3}$, which would result in only a handful of objects within our sample.

\subsection{Better Data: A Look Ahead}

Finally, we reiterate the need for deeper proper motion surveys in order to constrain the properties of the Galactic halo population. Future surveys like the Large Synoptic Survey Telescope, \textit{WFIRST}, \textit{Euclid} \citep{Euclid}, and \textit{CASTOR} \citep{Castor} will usher in a new era in deep wide-field surveys that will uncover thousands of halo white dwarf candidates. 

In order to emphasize this point, we have run a simulation with the mean parameters acquired as part of our fit to the depth of CFIS-\textit{u} (\textit{u} = 24.2). The resulting color-color diagram and RPMD can be seen in Figure \ref{fig:future}, where the right-hand panel shows the simulation with the CFIS magnitude limit and the left-hand panel is the same as Figure \ref{fig:result}. The resulting simulation with the CFIS magnitude limit returns half an order of magnitude more white dwarfs, and the disparity is particularly noticeable in the thick disk and halo populations. Specifically, we expect that our sample of halo objects would increase by a factor of 5$-$10, and our thick disk sample by a factor of three at the depth of the CFIS observations. Given their importance in understanding Galactic evolution, in particular the age of the halo via the white dwarf luminosity function, upcoming surveys such as LSST and \textit{WFIRST} will provide an unprecedented look into the evolution of the Galactic inner halo via its white dwarf population.

\section{Summary}
\label{sec:Conclusion}

In this paper, we have presented a newly developed white dwarf population synthesis code and used it to determine the star formation history of the Galactic thin disk, thick disk, and stellar halo using new exquisite \textit{u}-band data from CFIS. Our population synthesis model takes MW geometry and extinction, as well as survey parameters such as completeness, into account in order to return a mock catalog of white dwarfs in the specified photometric bands. We use data from the Canada-France Imaging Survey (CFIS), Pan-STARRS 1 (PS1) and \textit{Gaia}  DR2 to compare to our resulting simulations with a given star formation history, which we have parameterized using skewed Gaussian functions. This sample was shown to be local to the solar neighborhood, with a median distance of 388\,pc. Our main results are the following:

\begin{itemize}
 
  \item The star formation history of the Milky Way disk begins (11.3 $\pm$ 0.5)\,Gyr ago with the formation of the thick disk. The thick disk formed stars for 3\,Gyr and reached a maximum of (8.8 $\pm$ 1.4)\,M$_{\odot}$yr$^{-1}$ at (9.8 $\pm$ 0.3)\,Gyr. 
  
  \item The thick disk peak was followed by a decline in the star formation for 1\,Gyr before the thin disk began forming stars at a roughly constant rate of (3.5 $\pm$ 0.3)\,M$_{\odot}$yr$^{-1}$ for the past 8\,Gyr. The maximum star formation rate of the thin disk occurred at (8.4 $\pm$ 0.3)\,Gyr.  
  
  \item Although the star formation shape parameters for the stellar halo are relatively poorly constrained, we find a mean age for the inner halo of 10.9$^{+1.3}_{-1.1}$\,Gyr.
  
  \item Studying the residuals reveals variations from a unimodal function, with an increase by as much as 50\% at (3.3 $\pm$ 1.8)\,Gyr and a decrease of 30\% at (5.8 $\pm$ 1.1)\,Gyr.

  \item The resulting mass of the thin disk, thick disk, and stellar halo was found to be (3.4 $\pm$ 0.6) $\times$ 10$^{10}$\,M$_{\odot}$.
  
  \item The white dwarf space densities were found to be (4.8 $\pm$ 0.4) $\times$ 10$^{-3}$\,pc$^{-3}$ for the thin disk, (1.0 $\pm$ 0.2) $\times$ 10$^{-3}$\,pc$^{-3}$ for the thick disk, and (6.3 $\pm$ 2.4) $\times$ 10$^{-6}$\,\,pc$^{-3}$ for the halo. The fractional contributions of the thin disk, thick disk, and halo to the local white dwarf population were found to be (83 $\pm$ 5)\,\%, (17 $\pm$ 3)\,\%, and (0.11 $\pm$ 0.05)\,\% respectively.
  
  \item We find a He-fraction of (21 $\pm$ 3)\,\%. This is consistent with previous studies of the solar neighborhood.

\end{itemize}

Given the \textit{Gaia}-imposed magnitude limit of our survey, our dataset is dominated by disk stars in the solar neighborhood. Future deep proper motion surveys will be needed to increase the sample of halo white dwarfs in order to properly constrain its formation parameters. Upcoming surveys, like Pan-STARRS1 DR2, the Large Synoptic Survey Telescope,  as well as \textit{WFIRST}, \textit{Euclid}, and \textit{CASTOR} if they incorporate proper motions will open a new window into the formation of our MW at its earliest times.

\bigskip
\bigskip

\acknowledgments
We would like to thank the referee and all members of the Canada-France Imaging Survey for their insightful comments and contributions to this project.

This work was supported in part by the Natural Sciences and Engineering Research Council of Canada (NSERC). ES gratefully acknowledges funding by the Emmy Noether program from the Deutsche Forschungsgemeinschaft (DFG). NFM gratefully acknowledges support from the French National Research Agency (ANR) funded project ``Pristine'' (ANR-18-CE31-0017) along with funding from CNRS/INSU through the Programme National Galaxies et Cosmologie.

This work is based on data obtained as part of the Canada-France Imaging Survey, a CFHT large program of the National Research Council of Canada and the French Centre National de la Recherche Scientifique. Based on observations obtained with MegaPrime/MegaCam, a joint project of CFHT and CEA Saclay, at the Canada–France–Hawaii Telescope (CFHT) that is operated by the National Research Council (NRC) of Canada, the Institut National des Science de l’Univers (INSU) of the Centre National de la Recherche Scientifique (CNRS) of France, and the University of Hawaii. This research used the facilities of the Canadian Astronomy Data Centre operated by the National Research Council of Canada with the support of the Canadian Space Agency. 

The Pan-STARRS1 Surveys (PS1) and the PS1 public science archive have been made possible through contributions by the Institute for Astronomy, the University of Hawaii, the Pan-STARRS Project Office, the Max-Planck Society and its participating institutes, the Max Planck Institute for Astronomy, Heidelberg and the Max Planck Institute for Extraterrestrial Physics, Garching, The Johns Hopkins University, Durham University, the University of Edinburgh, the Queen's University Belfast, the Harvard-Smithsonian Center for Astrophysics, the Las Cumbres Observatory Global Telescope Network Incorporated, the National Central University of Taiwan, the Space Telescope Science Institute, the National Aeronautics and Space Administration under Grant No. NNX08AR22G issued through the Planetary Science Division of the NASA Science Mission Directorate, the National Science Foundation Grant No. AST-1238877, the University of Maryland, Eotvos Lorand University (ELTE), the Los Alamos National Laboratory, and the Gordon and Betty Moore Foundation.

This work has made use of data from the European Space Agency (ESA) mission
{\it Gaia} (\url{https://www.cosmos.esa.int/gaia}), processed by the {\it Gaia}
Data Processing and Analysis Consortium (DPAC,
\url{https://www.cosmos.esa.int/web/gaia/dpac/consortium}). Funding for the DPAC
has been provided by national institutions, in particular the institutions
participating in the {\it Gaia} Multilateral Agreement.

\end{document}